\documentclass[aps,prb,twocolumn]{revtex4}
\usepackage{graphicx}
\usepackage{amssymb}
\usepackage{amsfonts}
\usepackage{amsmath}
\usepackage{bm}
\usepackage{color}
\usepackage{pstricks}
\usepackage{pst-node}

\renewcommand\Re{\operatorname{\mathfrak{Re}}}
\renewcommand\Im{\operatorname{\mathfrak{Im}}}
\DeclareMathOperator{\const}{const}
\renewcommand{\vec}[1]{\mathbf{#1}}

\newcommand{\der}{\text{d}}
\newcommand{\bis}{{\prime\hspace{-0.1ex}\prime}}
\newcommand{\msms}{\scriptscriptstyle{--}}
\newcommand{\psms}{\scriptscriptstyle{+-}}
\newcommand{\msps}{\scriptscriptstyle{-+}}
\newcommand{\psps}{\scriptscriptstyle{++}}

\newcommand{\Xiso}{{\ensuremath{\Xi_{\text{so}}}}}

\newcommand{\Dnor}{{\ensuremath{D_{\text{nor}}}}}
\newcommand{\pd}{\ensuremath{p\text{-}d}}
\newcommand{\kp}{\ensuremath{k\cdot p}}
\newcommand{\etal}{\textit{et al.}}

\begin{document}

\title{Theory of spin waves in ferromagnetic (Ga,Mn)As}

\author{Agnieszka Werpachowska}
\affiliation{Institute of Physics, Polish Academy of Sciences, al.~Lotnik\'ow 32/46, PL-02-668 Warsaw, Poland}

\author{Tomasz Dietl}
\affiliation{Institute of Physics, Polish Academy of Sciences, al.~Lotnik\'ow 32/46, PL-02-668 Warsaw, Poland}
\affiliation{Institute of Theoretical Physics, University of Warsaw, PL-00-681 Warsaw, Poland}

\begin{abstract}
The collective behavior of spins in a dilute magnetic semiconductor is determined by their mutual interactions and influenced by the underlying crystal structure. Hence, we begin with the atomic quantum-mechanical description of this system using the proposed variational-perturbation calculus, and then turn to the emerging macroscopic picture employing phenomenological constants.
Within this framework we study spin waves and exchange stiffness in the $\pd$ Zener model of (Ga,Mn)As, its thin layers and bulk crystals described by the $spds^\ast$ tight-binding approximation. Analyzing the anisotropic part of exchange, we find that the Dzyaloshinskii-Moriya interaction may lead to a cycloidal spin arrangement and uniaxial in-plane anisotropy of diagonal directions in thin layers, resulting in a surface-like anisotropy in thicker films. We also derive and discuss the spin-wave contribution to magnetization and Curie temperature. Our theory reconstructs the values of stiffness determined from the temperature dependence of magnetization, but reproduces only partly those obtained from analyzing precession modes in (Ga,Mn)As thin films.
\end{abstract}


\maketitle


\section{Introduction}
\label{sec:section1}

Dilute magnetic semiconductors, such as (Ga,Mn)As, form a class of materials with many outstanding properties\cite{Matsukura:2002_B, Jungwirth:2006_RMP} and functionalities.\cite{Dietl:2007_IEEE} A number of these result from the complex structure of the valence band, which hosts holes mediating magnetic order between spins localized on transition metal impurities. The strong force which lies behind this mechanism is the carrier-mediated exchange interaction. It gives the low-lying energy states of such a system the character of spin waves, which contribute to both its equilibrium (e.g.~spontaneous magnetization and Curie temperature) and nonequilibrium magnetic properties (ferromagnetic resonance and relaxation).\cite{Farle:1998_RPP} At the same time, it allows us to replace the atomic quantum-mechanical description of the system with the classical, continuous micromagnetic theory, where it appears in form of exchange stiffness.\cite{Skomski:2008_B}

Quite generally, micromagnetic properties of any ferromagnet are determined by the magnitudes of the exchange stiffness and magnetocrystalline anisotropy.\cite{Kittel:1987_B} The first describes the exchange energy associated with nonuniform distributions of local directions of magnetization. The other is the energy needed to change the total magnetization direction with respect to the crystal axes, which involves the competing crystal-field and spin-orbit interactions. While the main part of the exchange energy is isotropic, a consequence of its electrostatic origin, the relativistic spin-orbit coupling can create its small anisotropy, namely the dependence on the crystalline orientation of magnetization. In zinc-blende bulk crystals and thin layers with broken space inversion symmetries, it leads to many interesting effects, related to the anisotropic\cite{Timm:2005_PRB} and Dzyaloshinskii-Moriya\cite{Dzyaloshinskii:1958_JPCS, Moriya:1960_PR} exchange, which have not been hitherto studied in (Ga,Mn)As. Since the total spin is no longer conserved, they are indispensable when considering any usage of spins for information storage and processing.

The spin-wave spectrum and the isotropic exchange stiffness in bulk (Ga,Mn)As were computed by K\"onig, Jungwirth and MacDonald,\cite{Konig:2001_PRB} and by Brey and G\'omez-Santos\cite{Brey:2003_PRB} within the $\pd$ Zener model employing the six-band $\kp$ Hamiltonian,\cite{Dietl:2000_S, Dietl:2001_PRB, Abolfath:2001_PRB} which neglects the inversion asymmetry specific to the zinc-blende lattice. It was found that the actual magnitude of the exchange stiffness is much greater when one takes into account the complex structure of the valence band, as compared to the case of a simple parabolic band.\cite{Konig:2001_PRB, Brey:2003_PRB} This, as well as the highly anisotropic Fermi surface, were shown to explain\cite{Brey:2003_PRB, Timm:2005_PRB} why the mean-field approximation\cite{Dietl:2000_S} is so accurate in (Ga,Mn)As. It was also found that the values of exchange and anisotropy energies obtained within the same formalism describe quantitatively\cite{Dietl:2001_PRBb} the width of stripe domains in films with perpendicular magnetic anisotropy.

More recently, Bouzerar\cite{Bouzerar:2007_EPL} employed a self-consistent local random-phase approximation in order to evaluate the spectral density of spin-wave excitations in Ga$_{1-x}$Mn$_{x}$As. The magnitudes of spin-spin exchange integrals $J(r)$ were obtained from first principles computations within the local spin-density approximation (LSDA) and tight-binding linear muffin-tin orbital approach neglecting the spin-orbit interaction. The theory allows to treat disorder and thermal fluctuations, and shows that the calculated spectral density has well-defined maxima up to about one half of the relevant Debye wave vector $q_{\text{D}} = (24x/\pi)^{1/3}\pi/a$, where $a$ is the lattice constant. This made it possible to determine the spin-wave dispersion $\omega(\vec{q})$ in the range $0<q \lesssim q_{\text{D}}/2$, from which the magnitude of spin-wave stiffness was obtained.\cite{Bouzerar:2007_EPL}

Experimentally, Potashnik \etal\cite{Potashnik:2002_PRB} analyzed the temperature dependence of magnetization in a series of Ga$_{1-x}$Mn$_x$As samples, which provided the values of spin-wave stiffness from the $T^{3/2}$ Bloch law. In later experiments, the stiffness was determined by examining spin precession modes excited by optical pulses\cite{Wang:2007_PRB} and under ferromagnetic resonance conditions.\cite{Zhou:2007_IEEE, Liu:2007_PRB, Bihler:2009_PRB} The values obtained for some films with thickness greater than 120~nm, either as-grown\cite{Wang:2007_PRB, Bihler:2009_PRB} or annealed,\cite{Wang:2007_PRB} are in good agreement with those predicted by Bouzerar.\cite{Bouzerar:2007_EPL} However, the values for thinner films\cite{Wang:2007_PRB, Bihler:2009_PRB} or another series of annealed samples\cite{Zhou:2007_IEEE, Liu:2007_PRB} were about three times smaller. Similarly small magnitudes of spin-wave stiffness were found by analyzing the domain structure of annealed Ga$_{0.93}$Mn$_{0.07}$As.\cite{Gourdon:2007_PRB} The experimental works\cite{Wang:2007_PRB, Zhou:2007_IEEE, Liu:2007_PRB, Bihler:2009_PRB} demonstrate that spectral positions of spin-wave resonances are strongly affected by the character of spin pinning at the sample borders. The same effect is caused by magnetic anisotropy changes along the growth direction, particularly strong at the film interface and surface.\cite{Bihler:2009_PRB} Therefore, theoretical predictions concerning spin-wave excitations in both thick and thin layers may provide a useful guide to better understanding of magnetization dynamics in real samples.

In this paper we investigate spin waves and related micromagnetic constants in ferromagnetic (Ga,Mn)As, as described by the $\pd$ Zener model.\cite{Dietl:2000_S} The validity of this model is supported by photoemission experiments\cite{Rader:2004_PRB} and {\em ab initio} computations, in which the inaccuracies of the LSDA are partly reduced by self-interaction.\cite{Schulthess:2005_NM} Furthermore, the magnitude of low-temperature quantum corrections to conductivity indicates that in the concentration range relevant to ferromagnetism, the density of states assumes values expected for valence band holes.\cite{Dietl:2008_JPSJ, Neumaier:2009_PRL} The carrier band is formed from $p$-type-like states of the GaAs semiconductor structure, which are mostly built from the anion $4p$ orbitals, but has a rather large $d$ component as the result of a strong $s\pd$ interaction. We describe it by the $spds^\ast$ tight-binding approximation for thin layers and, by applying periodic boundary conditions, for bulk crystals.\cite{Jancu:1998_PRB, Strahberger:2000_PRB, Sankowski:2007_PRB, Oszwaldowski:2006_PRB} Our analysis of the anomalous Hall effect in (Ga,Mn)As in Refs.~\onlinecite{Werpachowska:2010_PRB} and \onlinecite{Chiba:2010_PRL} provides a thorough comparison of this method with other band structure models.\cite{Werpachowska:2010_PRB} One of its advantages is that it captures the inversion asymmetry of the zinc-blende lattice, which produces the Dresselhaus spin splitting of the conduction band,\cite{Dresselhaus:1955_PR} and additionally the structure inversion asymmetry in thin layers, which creates the Bychkov-Rashba spin splitting.\cite{Bychkov:1984_JPh} The growth-induced biaxial strain is included by changing the atoms' arrangement, according to the strain tensor values: $\varepsilon_{xx} = \varepsilon_{yy} = \delta a/a$ and $\varepsilon_{zz} = -2\, c_{12}/c_{11}\, \varepsilon_{xx}$, where $\delta a$ is the strain-induced change of the lattice constant, and $c_{12}/c_{11} = 0.453$ is the ratio of elastic moduli. Also, the on-site energies of the $d$ orbitals depend linearly on the strain tensor values.\cite{Jancu:1998_PRB} The $s\pd$ exchange coupling is modelled using the virtual crystal and mean-field approximations, while taking into account the appropriate weights of Ga and As orbitals in the wavefunctions close to the center of the Brillouin zone.\cite{PhysRevB.65.161203,Oszwaldowski:2006_PRB} In this way, spin polarization of individual Mn moments is replaced by a molecular field, which creates a $k$-dependent Zeeman-like splitting of the host bands. For heavy holes in the $\Gamma$ point it is given by $\Delta = x n_0 S \beta$, where $x$ is the fraction of cation site density $n_0$ substituted by Mn forming a spin $S=5/2$, and $\beta = -54$~meV\,nm$^3$ is the $\pd$ exchange integral.\cite{Dietl:2001_PRB} The validity of this approach can be questioned if the magnetic ions produce bound states,\cite{Dietl:2008_PRB} but the metallic character of the carrier states of interest here means that this does not occur due to many-body screening. At the same time we neglect the effect of this screening on potentials produced by magnetization fluctuations, which is justified as long as $\Delta$ is smaller than the Fermi energy, so that spin and charge density fluctuations are decoupled. A part of Mn atoms form unintentional defects such as interstitials (which can passivate single substitutional spins) and antisites. Both being double donors, they significantly lower the hole density, which can be partly remedied by removing the interstitials by post-growth annealing.

The paper is arranged as follows. In Sec.~\ref{sec:theoreticalmodel} we use the proposed variational-perturbation calculus to describe the system of lattice spins interacting via hole carriers, and its spin-wave excitations. In the micromagnetic theory, the system is described by the set of constants related to magnetocrystalline anisotropy and exchange energy. Section~\ref{sec:exchange} provides quantitative results on the spin-wave stiffness expressed as the dimensionless parameter $\Dnor$, comparing it between the $\kp$ and $spds^\ast$ tight-binding models. In Sec.~\ref{subsec:magntemp} we derive the spin-wave contribution to magnetization and Curie temperature. Section~\ref{sec:experimental} compares our theory to related experimental findings.\cite{Potashnik:2002_PRB, Wang:2007_PRB, Zhou:2007_IEEE, Liu:2007_PRB, Bihler:2009_PRB, Gourdon:2007_PRB} The last section of the paper contains the summary of our work.

\section{Theoretical model}
\label{sec:theoreticalmodel}

In this section, we demonstrate a physically transparent perturbation-variational method of treating the systems in question. We use it to find the effective Hamiltonian of lattice ions interacting through hole carriers, written in the \textit{interaction representation} of creation and annihilation operators. Next, we calculate the dispersion dependence of its low lying energy states by transforming the Hamiltonian to the \textit{spin-wave representation}. These two pictures are associated with different sets of phenomenological parameters describing the macroscopic system, whose properties we shall investigate later on.

\subsection{Microscopic picture}
\label{subsec:effhamiltonian}

We consider the ferromagnetic phase of a system consisting of $P$ carriers and $N$ magnetic lattice ions, described by the Hamiltonian $\mathcal H_0$ and coupled by the $\pd$ exchange interaction $\mathcal H'$,
\begin{equation}
\label{eq:Hamiltonian}
\mathcal H = \mathcal{H}_0 + \mathcal{H'} = \mathcal H_0 + \sum_{i=1}^P \sum_{j=1}^{N}  \beta I(\vec{r}_i - \vec{R}_j)\, \vec{s}_i \cdot \vec{S}_j \ ,
\end{equation}
where $\vec{s}_i$ and $\vec{S}_j$ are the $i$-th carrier's and $j$-th ion's spin operators, while $\vec{r}_i$ and $\vec{R}_j$ are their respective positions. The strength of the $\pd$ exchange interaction between these two spins is described by a smooth function $\beta I(\vec{r}_i - \vec{R}_j)$, localized around the $j$-th magnetic ion. In the absence of external fields, $\mathcal H_0$ depends on the carriers' degrees of freedom only.\cite{Dietl:2001_PRB}

The dynamics of magnetic ions coupled to the system of hole carriers requires a self-consistent description, which takes into account how the holes react to the ions' magnetization changes. Therefore, we use the L\"{o}wdin perturbation method specifically adapted for multiparticle Hamiltonians,\cite{Loewdin:1951_JCP, Thijssen:2007_B, Ziener:2004_PRB, Werpachowska:2006_MS} to derive an effective Hamiltonian $\mathcal{H}^\text{eff}$ for ions only.

We choose the multiparticle basis states of $\mathcal H$ as $M \otimes \Gamma$. The ion part $M$ is an eigenstate of the unperturbed ions-holes system~\eqref{eq:Hamiltonian}, and the hole part $\Gamma$ is a Slater determinant of eigenstates of the one-particle hole Hamiltonian 
\begin{equation}
\label{eq:H1part}
h = h_0 + \Delta s^z\ ,
\end{equation}
where $h_0$ describes the host band structure and $\Delta$ is the spin splitting induced by polarized lattice ions.

The L\"{o}wdin calculus consists in dividing the multiparticle basis states into two subsets, $A$ and $B$,
\begin{equation}
\mathcal{H} = \begin{pmatrix}
\mathcal H_{AA} & \mathcal H_{AB} \\
\mathcal H_{BA} & \mathcal H_{BB}
\end{pmatrix} \ .
\end{equation}
Set $A$ contains all states $M \otimes \Gamma_0$, where $\Gamma_0$ is the $P$-particle ground state of $h$. Set $B$ contains all the remaining states, in which at least one hole is excited above the Fermi level. We construct the effective Hamiltonian for the states from set $A$ only, adding their coupling with set $B$ as the second order perturbation,
\begin{equation}
\label{eq:Heff}
\mathcal{H}^\text{eff}_{nn'} = (\mathcal{H}_0)_{nn'} + \mathcal H_{nn'}' + \sum_{n'' \in B} \frac{\mathcal H_{nn''} \mathcal H_{n''n'}}{E - \mathcal{H}_{n''n''}} \ ,
\end{equation}
where $n, n' \in A$. The term $(\mathcal{H}_0)_{nn'}$ is independent of the ion configurations, so we set it to zero for simplicity. Thus, the effective Hamiltonian $\mathcal{H}^\text{eff}$ depends only on the ion degrees of freedom but, thanks to the L\"{o}wdin method, takes hole excitations into account and can be used to calculate the spin-wave dispersion in a self-consistent manner.

The variational part of our method consists in searching for the energy $E$ in the range where we expect to find the lowest eigenenergies of $\mathcal H$, which are the ones that we are interested in. For a known average spin splitting $\Delta$, we can set $E$ to the total energy of the hole multiparticle state $\Gamma_0$, $E_{\Gamma_0} = \sum_{(\vec{k}, m) \in \Gamma_0} E_{\vec{k}, m}$, where $E_{\vec{k}, m}$ is the energy of the \mbox{$m$-th} band with wave vector $\vec{k}$ of Hamiltonian $h$, and the sum goes over all occupied eigenstates $(\vec{k}, m)$ in $\Gamma_0$. The states $n''$ are of the form $M'' \otimes \Gamma''$, $\Gamma'' \neq \Gamma_0$. To simplify the sum over $n''$, we approximate the diagonal matrix element $\mathcal{H}_{n''n''}$, which depends on both $M$ and $\Gamma''$, by the total energy of the multiparticle hole state $\Gamma''$, $E_{\Gamma''} = \sum_{(\vec{k}, m) \in \Gamma''} E_{\vec{k}, m}$. It describes the interaction of $\Gamma''$ with the average configuration of the ions' spins corresponding to the spin splitting $\Delta$. We can thus write the Hamiltonian~\eqref{eq:Heff} in the following form:
\begin{equation}
\label{eq:Heff2}
\mathcal{H}^\text{eff}_{nn'} = \mathcal H_{nn'}' + \sum_{M''} \sum_{\Gamma'' \neq \Gamma_0} \frac{\mathcal H_{nn''} \mathcal H_{n''n'}}{ E_{\Gamma_0} - E_{\Gamma''} } \ , \ n, n' \in A\ .
\end{equation}
The factor $\mathcal H_{nn''} \mathcal H_{n''n'}$ under the sum can be written as $\langle M \otimes \Gamma_0 | \mathcal{H} | \Gamma'' \otimes M'' \rangle \langle M'' \otimes \Gamma'' | \mathcal{H} | \Gamma_0 \otimes M' \rangle$, where $n = M \otimes \Gamma_0$ and $n' = M' \otimes \Gamma_0$. Since the denominator in Eq.~\eqref{eq:Heff2} is independent of $M''$, summing over $M''$ is equivalent to inserting an identity operator, which allows us to write the last term as
\[
\sum_{\Gamma'' \neq \Gamma_0} \frac{ \langle M \otimes \Gamma_0 | \mathcal{H} | \Gamma''  \rangle \langle \Gamma'' | \mathcal{H} | \Gamma_0 \otimes M' \rangle }{ E_{\Gamma_0} - E_{\Gamma''} } \ .
\]
We can thus treat $\mathcal{H}^\text{eff}$ as a Hamiltonian acting on ion states only,
\begin{equation}
\begin{split}
\label{eq:Heff-ions}
&\mathcal{H}^\text{eff}_{MM'} = \langle M \otimes \Gamma_0 | \mathcal{H}' | \Gamma_0 \otimes M' \rangle \\ &\ + \sum_{\Gamma'' \neq \Gamma_0} \frac{ \langle M \otimes \Gamma_0 | \mathcal{H} | \Gamma''  \rangle \langle \Gamma'' | \mathcal{H} | \Gamma_0 \otimes M' \rangle }{ E_{\Gamma_0} - E_{\Gamma''} } \ .
\end{split}
\end{equation}

Since the $\pd$ exchange term in $\mathcal{H}$, which produces the extra-diagonal matrix element $\langle M \otimes \Gamma_0 | \mathcal{H} | \Gamma'' \otimes M'' \rangle$, is the interaction of a single hole with an ion, the only $\Gamma''$ states which have a non-zero contribution to the sum over $\Gamma''$ in Eq.~\eqref{eq:Heff-ions} are those which are created from $\Gamma_0$ by just one excitation, $(\vec{k},m) \rightarrow (\vec{k}',m')$. Hence, we have $E_{\Gamma_0} - E_{\Gamma''} = E_{\vec{k}, m} - E_{\vec{k}',m'}$ and Hamiltonian~\eqref{eq:Heff-ions} can be written as
\begin{equation}
\label{eq:Heff-ions2}
\begin{split}
\mathcal{H}^\text{eff}_{MM'} &= \langle M \otimes \Gamma_0 | \mathcal{H}' | \Gamma_0 \otimes M' \rangle
\\&+ \sum_{\vec{k},\vec{k}'} \sum_{m,m'} \frac{f_{\vec{k},m} (1 - f_{\vec{k}',m'})}{ E_{\vec{k}, m} - E_{\vec{k}',m'} }
\\&\quad\times\langle M \otimes \Gamma_0 | \mathcal{H} | \Gamma''  \rangle \langle \Gamma'' | \mathcal{H} | \Gamma_0 \otimes M' \rangle\ ,
\end{split}
\end{equation}
where $f_{\vec{k},m}$ is the Fermi-Dirac distribution coefficient. The fraction in the above sum looks dangerous, as it may diverge in the presence of the energy bands' crossings, which would make our perturbation calculus invalid. However, the effective Hamiltonian for ions depends on the average of these factors, and will be shown immune to this problem.

We write Hamiltonian~\eqref{eq:Heff-ions} using ion spin operators and integrate out the hole degrees of freedom:
\begin{equation}
\label{eq:heisenberg}
\mathcal{H}^\text{eff} = \sum_{\sigma} \sum_{j=1}^{N} H_j^\sigma S_j^{\sigma} + \sum_{\sigma\sigma'} \sum_{j=1}^{N} \sum_{j'=1}^{N} H_{jj'}^{\sigma\sigma'} S_j^{\sigma} S_{j'}^{\sigma'}\ .
\end{equation}
The coefficients $H_j^\sigma$ and $H_{jj'}^{\sigma\sigma'}$ are given by
\begin{equation}
\label{eq:hjjss}
\begin{split}
H_{jj'}^{\sigma\sigma'} &= \frac{\beta^2}{V^2} \sum_{\vec{k}\vec{k'}} \sum_{mm'} \frac{f_{\vec{k},m} (1 - f_{\vec{k'},m'})}{E_{\vec{k'},m'} - E_{\vec{k},m}} \\
&\quad\times e^{i(\vec{k'}-\vec{k})\cdot(\vec{R}_j - \vec{R}_{j'})} s^{\sigma}_{\vec{k}m\vec{k}'m'} s^{\sigma'}_{\vec{k'}m'\vec{k}m}\ ,\\
\end{split}
\end{equation}
where due to the condition $\Gamma'' \neq \Gamma_0$ in Eq.~\eqref{eq:Heff2}, for $\vec k = \vec k'$ the summation goes over $m \neq m'$, and
\begin{equation}
\label{eq:hjs}
\begin{split}
H_j^\sigma &= \frac{\beta}{V} \sum_\vec{k} \sum_m f_{\vec{k},m} s^{\sigma}_{\vec{k}m\vec{k}m} \\
&+ \frac{\Delta \beta}{V} \sum_\vec{k} \sum_{m \neq m'} \frac{f_{\vec{k},m} (1 - f_{\vec{k'},m'})}{E_{\vec{k},m'} - E_{\vec{k},m}} \\
&\quad\times( s^\sigma_{\vec{k}m\vec{k}m'} s^z_{\vec{k}m'\vec{k}m} + s^\sigma_{\vec{k}m'\vec{k}m} s^z_{\vec{k}m\vec{k}m'} ) \ ,
\end{split}
\end{equation}
where $s^\sigma_{\vec{k}m\vec{k}'m'} = \langle u_{\vec{k},m} | \hat{s}^\sigma | u_{\vec{k'},m'} \rangle$ for $\sigma = +,-,z$ and $[\hat s^+, \hat s^-] = \hat s^z$ by convention. To obtain the above expressions, we substituted the hole-only part of Hamiltonian $\mathcal{H}_0$~\eqref{eq:Hamiltonian} by the sum of $P$ one-particle Hamiltonians $h_0$ from Eq.~\eqref{eq:H1part}. We also used the formula $\langle \psi_{\vec{k}, m} | h_0 | \psi_{\vec{k}',m'} \rangle = \delta_{\vec{k},\vec{k}'} (\delta_{m,m'} E_{\vec{k}, m} - \Delta s^z_{\vec{k}m\vec{k}'m'})$. It arises from the fact that the Bloch states $\psi_{\vec{k}, m} = e^{i \vec{k} \cdot \vec{r}} u_{\vec{k}, m}$ building the multiparticle hole states $\Gamma$ are one-particle eigenstates of Hamiltonian $h$~\eqref{eq:H1part}, which includes spin splitting $\Delta$. Assuming that $e^{i \vec{k} \cdot \vec{r}}$ is a slowly-varying function of $\vec r$ and $I(\vec r - \vec R_j)$ is constant within the unit cell around the $j$-th ion and vanishes outside of it, we obtained $\langle \psi_{\vec{k},m} | \hat{s}^\sigma I(\vec r - \vec R_j)| \psi_{\vec{k'},m'} \rangle = \frac{\beta}{V}\, e^{i(\vec k' - \vec k) \vec R_j} s^\sigma_{\vec{k}m\vec{k}'m'}$, where $V$ is the crystal volume.

The second term of $\mathcal{H}^\text{eff}$~\eqref{eq:heisenberg} describes the exchange interaction between the lattice ions. In the presence of the spin-orbit coupling it has an antisymmetric part in form of the Dzyaloshinskii-Moriya interaction,
\begin{equation}
	\label{eq:DMH}
	\sum_{\sigma^{\bis}} \sum_{jj'} i u_{jj'}^{\sigma^{\bis}} \sum_{\sigma\sigma'} \epsilon_{\sigma^{\bis}\sigma\sigma'} S_j^\sigma S_{j'}^{\sigma'}\ ,
\end{equation}
where $\epsilon_{\sigma^{\bis}\sigma\sigma'}$ is the antisymmetric Levi-Civita symbol with  $\epsilon_{{\psms} z} = 1$, while $\vec u_{ij}$ is a pseudovector and exists only in the systems with broken inversion symmetry,
\begin{equation*}
i u_{jj'}^{\sigma^\bis} = \frac{1}{2} \sum_{\sigma\sigma'} \epsilon_{{\sigma^\bis}\sigma\sigma'} \
H_{jj'}^{\sigma\sigma'}\ .
\end{equation*}

Using the L\"owdin perturbation-variational calculus, we have thus described the problem as a lattice spin system coupled by exchange interaction. The low-lying energy states of such systems are wavelike and can be modeled in the small oscillations approximation. For this purpose, we replace the spin operators with certain nonlinear functions of bosonic creation and annihilation operators $a_j^\dagger$ and $a_j$, carrying out the Holstein-Primakoff bosonization.\cite{Holstein:1940_PR} Next, we approximate these functions with their power expansions around the state of saturation magnetization:
\begin{equation}
\label{eq:spinoperators}
S_j^+ \approx \sqrt{S} a_j \ ,\ \ S_j^- \approx \sqrt{S} a_j^\dagger \ ,\ \ S_j^z = S - a_j^\dagger a_j\ ,
\end{equation}
leaving in the Hamiltonian the terms which are quadratic in creation and annihilation operators, as only these terms influence the dispersion relation. This approximation works very well in the long-wave limit, $a q \ll \pi$, as the neglected magnon-magnon interactions are proportional to $(a q)^4$.\cite{Kittel:1987_B} Furthermore, it justifies neglecting any short-range interactions between localized spins. At the same time, the continuous medium approximation becomes valid, which allows us to carry out the Fourier transform $a_{\vec{q}} = N^{-1/2} \sum_{j=1}^{N} e^{i \vec{q} \cdot \vec{R}_j} a_j$. After simple algebraic transformations, we arrive at the final form of the harmonic Hamiltonian,
\begin{equation}
\label{eq:Ha}
\begin{split}
\mathcal H^\text{eff} = &\sum_{\vec{q}} \Bigl[ \left( \Xi - \chi_{\vec{q}}^{\psms} \right) a_{\vec{q}}^\dagger a_{\vec{q}} \\
&- \frac{1}{2} \chi_{\vec{q}}^{\psps} a_{\vec{q}} a_{-\vec{q}} - \frac{1}{2} \chi_{\vec{q}}^{\msms} a_{\vec{q}}^\dagger a_{-\vec{q}}^\dagger \Bigr] \ .
\end{split}
\end{equation}
We call it the \textit{interaction} representation as it describes the perturbation of the ground state by the isotropic Coulomb interaction (first term) and by the spin-orbit interaction, coupling modes of different $\vec{q}$ (remaining terms). The spin susceptibility of the holes is given by
\begin{equation}
\begin{split}
\label{eq:chi}
\chi^{\sigma\sigma'}_\vec{q} = &-\frac{n S \beta^2}{V} \sum_{\vec{k}} \sum_{mm'} \frac{f_{\vec{k},m} - f_{\vec{k}+\vec{q},m'}}{E_{\vec{k},m} - E_{\vec{k}+\vec{q},m'}} \\
&\quad\times s^{\sigma}_{\vec{k}m(\vec{k}+\vec{q})m'} s^{\sigma'}_{(\vec{k}+\vec{q})m'\vec{k}m}\ ,
\end{split}
\end{equation}
where $n=N/V$ is the density of localized spins $S$ in the sample volume $V$ and $n S \beta = \Delta$. The presence of the energy denominator shows that $\chi^{\sigma\sigma'}_\vec{q}$ corresponds to the second-order part of the Hamiltonian~\eqref{eq:Heff-ions2}. As promised, vanishing of the denominator is not harmful, due to the de l'Hospital rule.

The formula~\eqref{eq:chi} implies that $\chi^{\psps}_\vec{q} = (\chi^{\msms}_\vec{q})^\ast$ is symmetric in $\vec q$, while $\chi^{\psms}_\vec{q}$, $\chi^{\msps}_\vec{q} \in \mathcal R$ inherit the symmetry of the $\psi_{\vec{k}, m}$ eigenstates. Hence, we expect the latter to be symmetric with respect to $\vec q$ for systems which preserve space inversion symmetry,\cite{Konig:2001_PRB} and otherwise for systems which do not.
The $\vec{q}$-independent term describes the interaction of a single magnetic ion with a molecular field arising from the intraband spin polarization of the carriers,
\begin{equation}
\label{eq:Xi}
\Xi = -\frac{\beta}{V} \sum_{\vec k} \sum_m f_{\vec k,m}\, s^z_{\vec km\vec km}\ .
\end{equation}
The corresponding term reflecting the interband polarization,
\begin{equation}
\label{eq:Xiso}
\Xiso = \frac{n S \beta^2}{V} \sum_{\vec{k}} \sum_{m\neq m'} \frac{f_{\vec{k},m} - f_{\vec{k},m'}}{E_{\vec{k},m'} - E_{\vec{k},m}} |s^z_{\vec{k}m\vec{k}m'}|^2\ ,
\end{equation}
arises from both $H_{jj'}^{\sigma\sigma'}$~\eqref{eq:hjjss} and the second part of $H_j^\sigma$ coefficient~\eqref{eq:hjs}, and cancels exactly in the full Hamiltonian $\mathcal H^\text{eff}$.

Hamiltonian~\eqref{eq:Ha} in the interaction representation describes the spin system in terms of circularly polarized plane waves (first term), which interact with each other and deform in time (remaining terms). We want to obtain the dispersion relation of independent, stable magnons. For this purpose, we diagonalize $\mathcal H^\text{eff}$ by the Bogoliubov transformation from $a_{\vec{q}}, a_{\vec{q}}^\dagger$ to $b_{\vec{q}}, b_{\vec{q}}^\dagger$ operators, keeping in mind that we deal with the system which breaks the space inversion symmetry. The final form of the effective Hamiltonian in the \textit{spin-wave} representation reads
\begin{equation}
\label{eq:Hbb}
\mathcal{H}^\text{eff} = -\sum_{\vec{q}} \omega_{\vec{q}} b_{\vec{q}}^{\dagger} b_{\vec{q}} \ ,
\end{equation}
where excitation modes are spin waves with dispersion
\begin{equation}
\label{eq:omega}
\omega_\vec{q} = \frac{\chi_{\text{\small{-}}\vec{q}}^{\psms} {\scriptstyle -} \chi_\vec{q}^{\psms}}{2} + \sqrt{ \frac{(2\Xi {\scriptstyle -} \chi_\vec{q}^{\psms} {\scriptstyle -} \chi_{\text{\small{-}}\vec{q}}^{\psms})^2}{4} - |\chi_\vec{q}^{\psps}|^2}\,.
\end{equation}
In the case of $\chi_\vec{q}^{\psms} = \chi_{-\vec{q}}^{\psms}$, fulfilled for the systems invariant under space inversion, the above formula simplifies to K\"{o}nig's~\textit{et al.}\cite{Konig:2001_PRB} solution,
\begin{equation}
\omega_{\vec{q}} = \sqrt{\left(\Xi - \chi_{\vec{q}}^{\psms}\right)^2 - |\chi_{\vec{q}}^{\psps}|^2}\ .
\end{equation}
Furthermore, neglecting the spin-orbit coupling, when $\chi^{\psps} = 0$, Bogoliubov transformation is unnecessary, and the effective Hamiltonian is already diagonalized by $a_{\vec{q}}, a_{\vec{q}}^\dagger$ operators. The negative sign of the Hamiltonian~\eqref{eq:Hbb} is a consequence of using the electronic convention to describe the hole-ion system.

Apart from spin waves, which are the eigenstates of the stationary Hamiltonian $\mathcal H^\text{eff}$, dynamic excitations of different physical origin may occur (e.g.\ Stoner spin-flips). They transfer a single carrier across the Fermi level to the state excited by the energy $\hbar\omega = E_{\vec{k},m} - E_{\vec{k}+\vec{q},m'}$ (see the denominator of Eq.~\eqref{eq:chi}, and may lead to the spin waves' damping. In the presence of the spin-orbit coupling, we find them likely to appear at very low energetic cost throughout the whole $\vec{q}$-vector range. On the other hand, within the mean-field approximation and well below the Curie temperature, substitutional and thermal disorder are characterized by a short correlation length $\xi$, which ensures a well-defined spin-wave excitation spectrum for $q < 2\pi/\xi$.

\begin{figure}
\centering
  \includegraphics[width=0.9\linewidth]{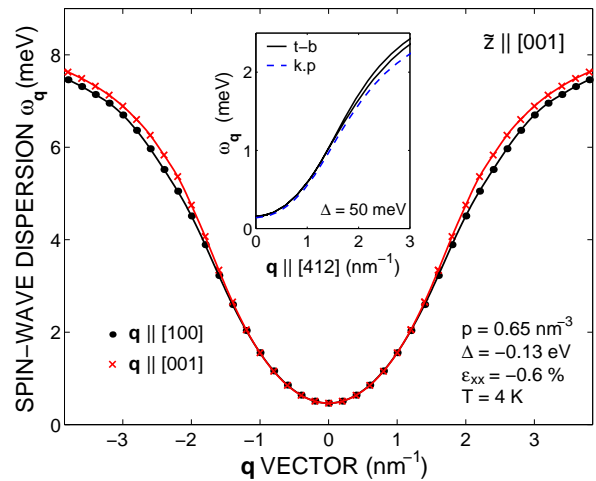}
\caption{(Color online) Dispersion dependence of spin waves propagating along $[100]$ (equivalent to $[010]$) and $[001]$ directions in bulk (Ga,Mn)As. Inset: the $\omega_\vec q$ asymmetry due to the bulk inversion asymmetry of GaAs lattice modeled by the $spds^\ast$ tight-binding method, as compared to the six-band $\kp$ model preserving the symmetry. Lower and upper solid lines denote the [412] and the opposite $[\bar4\bar1\bar2]$ propagation direction.}
\label{fig:figure1}
\end{figure}

Figure \ref{fig:figure1} presents a typical spin-wave dispersion spectrum $\omega_\vec{q}$~\eqref{eq:omega} in bulk (Ga,Mn)As calculated by the tight-binding computational scheme described in Sec.~I. The hole concentration $p$ equals $0.65$\,nm$^{-3}$ and the spin splitting $\Delta = -0.13$\,eV is applied along the easy axis $\tilde z$ fixed to the [001] direction by the biaxial strain $\varepsilon_{xx} = -0.6\%$.\cite{Dietl:2001_PRB} The latter causes a small anisotropy between $\omega_\vec{q}$ for spin waves propagating in the $[001]$ direction and in the $(001)$ plane. Furthermore, due to the lack of inversion symmetry in the unit-cell geometry, $\chi^{\psms}_{\vec{q}} \neq \chi^{\psms}_{-\vec{q}}$ in general. This means that, contrary to the results of the six-band $\kp$ model,\cite{Konig:2001_PRB} $\omega_{\vec{q}}$ can be asymmetric with respect to the sign of $\vec q$ or $\Delta$ (inset). (This effect was studied experimentally and theoretically in the context of dielectric susceptibility in another zinc-blende crystal, InSb.\cite{Chen:1985_PRB}) Related to the very small Dresselhaus spin splitting of the conduction band, the also small $\omega_{\vec{q}}$ asymmetry is observed for all but [100] and [111] $\vec q$ directions, and decreases with the growth of spin splitting. This result remains in line with our previous studies showing that static properties of (Ga,Mn)As like Curie temperature or anisotropies, to which $\omega_\vec{q}$ belongs, depend mainly on the properties of the carrier bands.\cite{Werpachowska:2010_PRB}

\begin{figure}
\centering
  \includegraphics[width=0.9\linewidth]{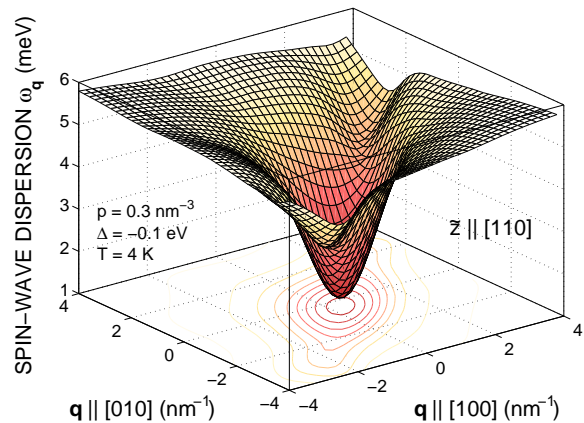}
\caption{(Color online) Dispersion dependence of spin waves in two (Ga,Mn)As monolayers. The minimum shift of in the $[1\bar10]$ direction to $\vec q_\text{min}=(0.43,-0.43)$\,nm$^{-1}$ can be observed.}
\label{fig:figure2}
\end{figure}

Figure \ref{fig:figure2} presents $\omega_\vec{q}$ in (Ga,Mn)As thin layer for $p=0.3$\,nm$^{-3}$ and $\Delta = -0.1$\,eV along the easy axis [110]. The host crystal consists of two unstrained infinite monolayers (Ga, As, Ga, As) grown in the [001] direction. The new effect, related to the structure asymmetry, is a shift of the dispersion minimum to a non-zero $\vec q_\text{min}$ value.

According to the above plots, $\omega_{\vec q}$ for small $\vec q$-vectors can be described by the following general formula:
\begin{equation}
\label{eq:omegamacro}
\omega_{\vec{q}} = D^{\mu\nu} q_\mu q_\nu - U^\mu q_\mu + \omega_0\ ,
\end{equation}
where indices $\mu, \nu = x,y,z$ denote spatial directions used in the Einstein sum convention. The $\mathbf D$ and $\mathbf U$ constants are the spin-wave stiffness tensor and the Dzyaloshinskii-Moriya coefficient, respectively, while $\omega_0$ is the spin-wave gap created by magnetocrystalline anisotropies. The higher order terms arising from the bulk inversion asymmetry can be skipped as negligibly small for considered $\vec q$-vectors.

Alternatively, we can expand the $\vec q$-dependent terms in $\omega_{\vec q}$, considering their properties implied by Eq.~\eqref{eq:chi},
\begin{equation}
	\label{eq:chiexp}
	\begin{split}
	&\chi^{\psms}_\vec{q} \approx \frac{2 g \mu_\text{B}}{M_S} (A^{\mu\nu} q_\mu q_\nu + \tilde U^\mu q_\mu) + \chi^{\psms}_{\vec{q}=0}\ ,\\
	&\chi^{\psps}_\vec{q} \approx \frac{2 g \mu_\text{B}}{M_S} T_{\psps}^{\mu\nu} q_\mu q_\nu + \chi^{\psps}_{\vec{q}=0}\ ,
	\end{split}
\end{equation}
where $M_S = g \mu_\text{B} n S$ is the saturation magnetization of the spin system. We obtain that, up to quadratic terms, the spin-wave stiffness in Eq.~\eqref{eq:omegamacro} depends only on the $\chi^{\psms}_\vec{q}$ term,
\begin{equation}
	\label{eq:coeff}
	\mathbf D =\frac{2 g \mu_\text{B}}{M_S} \mathbf A\ ,\ \text{while}\ \vec U = \frac{2 g \mu_\text{B}}{M_S} \tilde{\vec U}\ .
\end{equation}

The above constants are the subject of micromagnetics, which we shall investigate in the next section. It ignores the quantum nature of the atomic matter and uses classical physics in the limit of a continuous medium.
	
\subsection{Macroscopic picture}
\label{subsec:exchstiffness}

The atomic-scale effects investigated in the previous section lead to the wavelike behavior of the spin system. In micromagnetics, these spins are replaced by classical vectors with their slow-varying direction $\vec n(\vec r)$ described by the free energy functional:
\begin{equation}
\label{eq:EgeneralAK}
\begin{split}
E[\vec{n(\vec r)}] = \int \Bigl[ \sum_{j=1}^\infty \mathcal K_j^{\mu} n_{\mu}^{2j}
&+ \mathcal{A}_{\alpha\beta}^{\mu\nu} \partial_\mu n^\alpha \partial_\nu n^\beta \\
&+ U^\mu \epsilon_{\alpha\beta} n^\alpha \partial_\mu n^\beta \Bigr] \der^3\vec{r}\ ,
\end{split}
\end{equation}
where indices $\mu, \nu = x,y,z$ and $\alpha,\beta = \tilde x,\tilde y$ denote spatial and magnetization directions, respectively, and $\epsilon_{\alpha\beta}$ is the antisymmetric Levi-Civita symbol. The first term describes the anisotropy energy, which depends on the orientation of the magnetization with respect to the easy axis $\tilde z$. Consecutive orders of the magnetocrystalline anisotropy tensor $\boldsymbol{\mathcal{K}}$ in principal-axis representation are numbered by $j$. The next term is the symmetric exchange energy, where $\boldsymbol{\mathcal{A}}$ is the exchange stiffness tensor, with $\mathcal{A}_{\alpha\beta}^{\mu\nu} = \mathcal{A}_{\beta\alpha}^{\nu\mu}$. The antisymmetric part of exchange is expressed by the last, Dzyaloshinskii-Moriya term.\cite{Dzyaloshinskii:1958_JPCS, Moriya:1960_PR} Since we describe the magnetization fluctuations around the easy axis, we do not include the derivatives of $n^{\tilde z}$ in the sum, as they are of higher order.

The exchange stiffness $\boldsymbol{\mathcal{A}}$ can be split into two parts. The first one is isotropic in the magnetization direction, but can bear the anisotropy of the exchange interaction in space (e.g.\ caused by biaxial strain),
\begin{equation}
\label{eq:Amunu}
A^{\mu\nu} = \frac{1}{2} \left( \mathcal{A}_{xx}^{\mu\nu} + \mathcal{A}_{yy}^{\mu\nu} \right)\ .
\end{equation}
The remaining part is anisotropic with respect to the magnetization direction,
\begin{equation}
\label{eq:Tmunu}
T^{\mu\nu}_{\alpha\beta} = \mathcal{A}_{\alpha\beta}^{\mu\nu} - A^{\mu\nu} \delta_{\alpha\beta} \ .
\end{equation}

We define indices $\sigma,\sigma' = +,-$, referring to corresponding spin components, so that $n^\pm = (n^{\tilde x} \pm i n^{\tilde y}) / \sqrt{2}$ and $\mathcal{A}_{\alpha\beta}^{\mu\nu} \partial_\mu n^\alpha \partial_\nu n^\beta = \mathcal{A}_{\sigma\sigma'}^{\mu\nu} \partial_\mu n^\sigma \partial_\nu n^\sigma$. It follows from the tensors' definitions that $(T_{\psps}^{\mu\nu})^\ast = T_{\msms}^{\mu\nu}$ and $T_{\psms}^{\mu\nu} = 0$. We shall use the new notation to rewrite the exchange energy, including the Dzyaloshinskii-Moriya term, in the following form:
\begin{equation}
\label{eq:energy}
\begin{split}
E_\text{ex} = \int \Bigl[ &2 A^{\mu} \partial_\mu n^+ \partial_\mu n^- + T_{\sigma\sigma}^{\mu\nu} \partial_\mu n^\sigma \partial_\nu n^\sigma \\
&\ + iU^\mu ( n^+ \partial_\mu n^- - n^- \partial_\mu n^+) \Bigl] \der^3\vec{r} \ .
\end{split}
\end{equation}

In analogy to microscopic approach, we transform the exchange energy functional to the reciprocal space,
\begin{equation}
\label{eq:Eexbosonic}
\begin{split}
&E_\text{ex} = \frac{g \mu_\text{B}}{M_S} \sum_{\vec{q}} \Bigl[ \left(2 A^{\mu} q_\mu^2 + \tilde U^\mu q_\mu\right)\,a(\vec{q})^\dagger a(\vec{q}) +
\\&\quad\, T^{\mu\nu}_{\psps} q_\mu q_\nu\, a(\vec{q})^\dagger a(-\vec{q})^\dagger + T^{\mu\nu}_{\msms} q_\mu q_\nu\, a(\vec{q}) a(-\vec{q}) \Bigr] \ .
\end{split}
\end{equation}
We compare the above result to the effective Hamiltonian~\eqref{eq:Ha} in the interaction representation. It is clear that $\mathbf A$ (together with $\mathbf U$) and $\mathbf T$ correspond to coefficients of the $\chi^{\psms}_\vec{q}$ and $\chi^{\psps}_\vec{q}$ expansions in Eq.~\eqref{eq:chiexp}, respectively. Furthermore, we can identify the components of the above form with different physical mechanisms governing the spin behavior. The term involving $\mathbf A$ describes the energy of a circularly polarized spin wave, as it arises from the isotropic part of exchange interaction. Hence, we shall call it the \textit{isotropic} exchange stiffness tensor. The two terms involving $\mathbf T$ account for the anisotropic exchange, as we have chosen in Eq.~\eqref{eq:Tmunu}, induced by the spin-orbit coupling. Hence, we shall call it the \textit{relativistic} exchange stiffness tensor. Its non-zero elements imply that the tilting of an individual spin from the easy axis $\tilde z$ to different directions has different energetic cost. As a consequence, the polarization of the spin wave deforms and acquires an elliptical shape. The linear term characterized by the constant $\tilde{\mathbf U}$ represents the minimum shift of the spin-wave dispersion dependence observed in thin (Ga,Mn)As layers (Fig.~\ref{fig:figure2}), associated with the asymmetric exchange of Dzy\-alo\-shin\-skii-Moriya~\eqref{eq:DMH}. The energy $\omega_0$ in Eq.~\eqref{eq:omegamacro} is related to the anisotropy constant $\boldsymbol{\mathcal K}$ in the full free energy functional~\eqref{eq:EgeneralAK}.

We calculate the above tensors for the bulk (Ga,Mn)As from Fig.~\ref{fig:figure1} and two monolayers from Fig.~\ref{fig:figure2}. For this purpose, we fit the coefficients of the $\chi^{\psms}_\vec{q}$ and $\chi^{\psps}_\vec{q}$ expansions in Eq.~\eqref{eq:chiexp} on ca 1\,nm$^{-1}$ edge cube (or square) in $\vec q$-space, centered around zero. The obtained tensors describe the energy of spin waves polarized in the plane perpendicular to the easy axis $\tilde z$. If they propagate in this plane, we call them longitudinal waves. Transverse spin waves propagate along $\tilde z$.

\subsubsection{Bulk (Ga,Mn)As}
\label{sec:bulk}

The dispersion spectrum of spin waves propagating in bulk (Ga,Mn)As along two main crystal axes, $[100] \parallel x$ (equivalent to $[010]$) and $[001] \parallel z$, is presented in Fig.~\ref{fig:figure1}. The simulated system is biaxially strained, $\varepsilon_{xx}=-0.6\%$, with the hole concentration $p=0.65$\,nm$^{-3}$ and spin splitting $\Delta=-0.13$\,eV along the easy axis $\tilde z \parallel [001]$.

The energy cost of exciting a circularly polarized wave is given by the exchange stiffness tensor
\begin{equation}
\label{eq:Abulk}
\mathbf A =
\begin{pmatrix}
1.32 & 0 & 0\\
0 & 1.32 & 0\\
0 & 0 & 1.28
\end{pmatrix}\text{meV nm$^{-1}$}\,.
\end{equation}
It is expressed by a diagonal form with eigenvectors pointing along crystal axes, as the magnetization in the spins' ground state is uniform. The difference between its elements reflects the anisotropy of the exchange interaction in space (between the $xy$ plane and the growth direction $z$) caused by the biaxial strain.

\begin{figure}
\centering
  \includegraphics[width=0.8\linewidth]{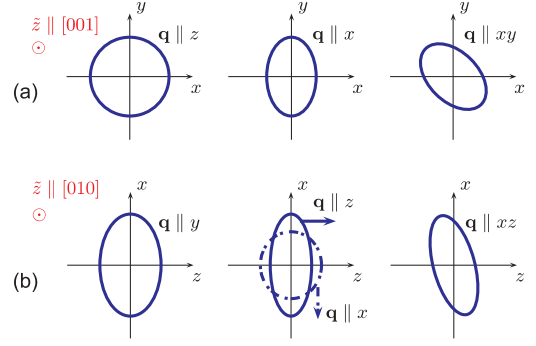}
\caption{(Color online) Spin-wave polarization (exaggerated for clarity), namely a shape traced out in a fixed plane by the spin vector rotating around the magnetization easy axis $\tilde z$, in (a) the bulk (Ga,Mn)As from Fig.~\ref{fig:figure1} with $\tilde z \parallel [001]$ and (b) after changing $\tilde z$ to [010].}
\label{fig:figure4}
\end{figure}

In the presence of the spin-orbit coupling, the circular polarization can deform into an ellipse. This polarization anisotropy is described by the relativistic $\mathbf T_{\psps}$ tensor, which depends on the mutual orientation of magnetization and spin-wave propagation directions. In our system with the easy axis $\tilde z$ along the spatial $z$ direction, it takes the following form:%
\begin{equation}
\label{eq:Tpp1}
\mathbf T_{\psps} =
\begin{pmatrix}
0.020 & -0.035\, i & 0\\
-0.035\, i & -0.020 & 0\\
0 & 0 & 0
\end{pmatrix} \text{meV nm$^{-1}$}\,.
\end{equation}
Its zero diagonal component means that the polarization of transverse spin waves is circular, while the non-zero elements indicate that longitudinal spin waves have elliptical polarization, with the shorter axis of the ellipse rotated to the $\vec q$ direction. The resulting polarizations are illustrated with Fig.~\ref{fig:figure4}\,a.

\begin{figure}
\centering
  \includegraphics[width=0.8\linewidth]{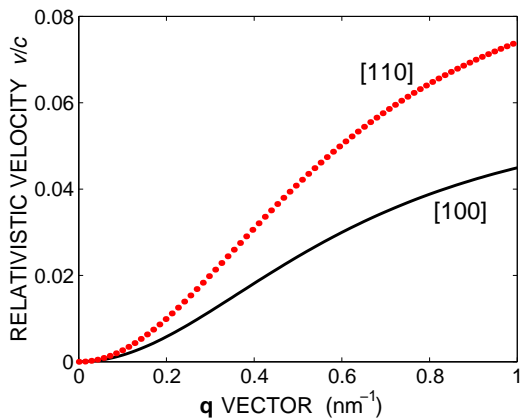}
\caption{(Color online) Relativistic velocity of spin waves propagating in two crystallographic directions: [110] (dotted line) and [100] (solid line), as a function of the $\vec q$-vector magnitude.}
\label{fig:figure3}
\end{figure}

For an arbitrary propagation direction, the shape of the spin-wave polarization can be calculated from the following polarization matrix:
\begin{equation}
\begin{split}
\label{eq:plrztnmtrx}
&\qquad\qquad\mathbf p = \begin{pmatrix}
p_{\tilde x\tilde x} & p_{\tilde x\tilde y}\\
p_{\tilde y\tilde x} & p_{\tilde y\tilde y}
\end{pmatrix}\ ,
\end{split}
\end{equation}
where $p_{\alpha\beta} = \mathcal A^{\mu\nu}_{\alpha\beta} q_\mu q_\nu$, or in more detail
\begin{equation}
\begin{split}
&p_{\tilde x\tilde x}=\vec q\, (\mathbf A + \Re{\mathbf T_{\psps}})\, \vec q^\mathrm{T} + \chi^{\psms}_0 + \Re\chi^{\psps}_0 - \Xi\ ,\\
&p_{\tilde y\tilde y}=\vec q\, (\mathbf A - \Re{\mathbf T_{\psps}})\, \vec q^\mathrm{T} + \chi^{\psms}_0 - \Re\chi^{\psps}_0 - \Xi\ ,\\
&p_{\tilde x\tilde y}=p_{\tilde y\tilde x}= -\Im\vec q\mathbf T^{\psps}\vec q^\mathrm{T}-\Im\chi^{\psps}_0\ .
\end{split}
\end{equation}
The ellipse is the solution of the equation
\begin{equation}
n^\alpha n^\beta p_{\alpha\beta} = \const\ ,
\end{equation}
where $n^\alpha$ and $n^\beta$ are the in-plane components of the magnetization unit vector $\vec n$. The ratio of its longer and shorter main axes, $a$ and $b$, can be derived from the eigenvalues of the polarization matrix~\eqref{eq:plrztnmtrx} as
\begin{equation}
a/b = \sqrt{\frac{p_{\tilde x\tilde x} + p_{\tilde y\tilde y} + \sqrt{(p_{\tilde x\tilde x} - p_{\tilde y\tilde y})^2 + 4 p_{\tilde x\tilde y}^2}}{p_{\tilde x\tilde x} + p_{\tilde y\tilde y} - \sqrt{(p_{\tilde x\tilde x} - p_{\tilde y\tilde y})^2 + 4 p_{\tilde x\tilde y}^2}}} \ .
\end{equation}
Kinetically, a longitudinal elliptical spin wave can be viewed as a circular one, which experiences the Lorentz contraction in the direction of motion, $b = a \sqrt{1-v^2/c^2}$, traveling with the velocity $v$, where $c$ is the speed of light. Figure~\ref{fig:figure3} presents this \textit{spin-wave relativistic velocity} in the relevant range of $\vec q$ vectors along [100] and [110] directions.

As already mentioned, $\mathbf T$ is induced by the spin-orbit coupling, which connects the symmetries of the lattice with spins. Thus, similarly to magnetocrystalline anisotropies, it depends on the magnetization direction with respect to the crystal axes. The $\mathbf T$ tensor calculated in the analyzed (Ga,Mn)As system with the easy axis $\tilde z$ changed to the $[010]$ direction would acquire the following form:
\begin{equation}
\begin{split}
\mathbf T_{\psps} =
&\begin{pmatrix}
-0.018 & 0 & 0.034\,i\\
0 & -0.033 & 0\\
0.034\,i & 0 & 0.007
\end{pmatrix} \text{meV nm$^{-1}$}\,,
\end{split}
\end{equation}
which reveals its dependence on the biaxial strain. (An appropriate rearrangement of the matrix elements gives the $\mathbf T_{\psps}$ tensor for $\tilde z$ along [100].) Now the longitudinal spin waves propagate in the $\tilde x \tilde y \parallel (010)$  plane and experience the `Lorentz contraction' as described above, while the transverse waves propagate along the new easy axis $\tilde z$. Their polarizations are additionally deformed by the potential of the strained crystal. The latter is stretched equally in the [100] and [010] directions and compressed in the [001] direction, and so are the polarizations. The magnitude of their deformations is given by the $\vec q$-independent terms of the polarization matrix $\mathbf p$~\eqref{eq:plrztnmtrx}. Figure~\ref{fig:figure4}\,b is illustrative of this effect.

Technically, the relativistic exchange might lead to macroscopic anisotropies in the system, if they did not average out for spin waves propagating in different directions. However, one can imagine a weak anisotropy arising from the described phenomena in asymmetrically shaped samples, where the largest number of similarly polarized modes can exist along the longest dimension. It would then be a candidate for an explanation of the weak uniaxial anisotropy of the [100] and [010] crystal axes observed in some (Ga,Mn)As samples.\cite{Pappert:2007_APL}

\subsubsection{(Ga,Mn)As layers}
\label{sec:layers}

In thin layers of (Ga,Mn)As (Fig.~\ref{fig:figure2}), we observe the minimum shift of the spin-wave dispersion to a non-zero $\vec q_\text{min}$ value, which was not present in bulk. It is a hallmark of the Dzyaloshinskii-Moriya asymmetric exchange~\eqref{eq:DMH}, arising from structure inversion symmetry breaking. The mechanism of this interaction is demonstrated in our numerical simulations for the two monolayers of (Ga,Mn)As from Fig.~\ref{fig:figure2}.

\begin{figure}
\centering
  \includegraphics[width=0.65\linewidth]{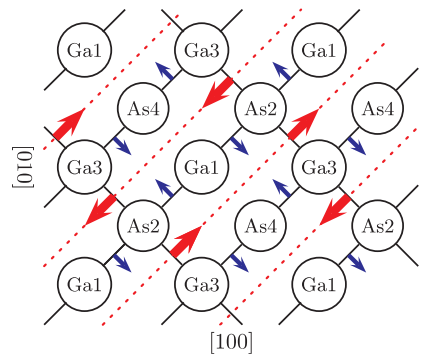}
\caption{(Color online) Two infinite monolayers of GaAs. The Ga and As sites are numbered by their elevation in the growth direction [001]: 1's are situated at $(\cdot,\cdot,0)$, 2 at $(\cdot,\cdot,\frac{1}{4}a)$, 3 at $(\cdot,\cdot,\frac{1}{2}a)$ and 4 at $(\cdot,\cdot,\frac{3}{4}a)$. The molecular field of magnetic Mn ions is introduced according to Sec.~\ref{sec:section1}. The strongest interactions between the nearest neighbors are marked by solid bonds. The created solid lines define mirror planes in the zinc-blende crystal, (110) and $(1\bar10)$. Since the analyzed structure has two-dimensional periodicity, the carrier momenta in the growth direction are quantized and spin waves propagate in-plane only.}
\label{fig:figure5}
\end{figure}

The structure of the modeled lattice is shown in Fig.~\ref{fig:figure5}. Thanks to the lack of inversion symmetry at the midpoint of each Ga-As bond, the exchange interaction of Dzyaloshinskii-Moriya is allowed, $\vec u_{jj'}\cdot \vec S_j\times\vec S_{j'}$. According to Moriya's rules,\cite{Moriya:1960_PR} the $\vec u_{ij}$ vector of each atom pair, indicated by an arrow at each bond, is perpendicular to their mirror plane. Its sense is always the same when we go from Ga to As atom. Since in our theoretical approach outlined in Sec.~\ref{subsec:effhamiltonian} we have used the Holstein-Primakoff transformation to describe spin waves as small fluctuations around the magnetization direction $\tilde z$, only the $u_{jj'}^{\tilde z}$ terms contribute to the spin-wave dispersion relation. For the above reasons, we expect the maximum effect of the Dzyaloshinskii-Moriya interaction in systems magnetized along the $\vec u$ vectors: perpendicular to $[110]$ or $[1\bar10]$. First, we consider the magnetization $\tilde z$ set along the $[110]$ direction and a spin wave propagating perpendicular to it, $\vec q \parallel [1\bar10]$. The dotted line is the wavefront, along which all spins must be in phase. Along this wavefront, the constant $\vec u$ vector (big arrows) tilts the spins perpendicular to itself and to each other to minimize the energy of the Dzyaloshinskii-Moriya interaction. The $\vec u_{jj'}$ vector is a function of the distance between the spins, $\vec R_j-\vec R_{j'}$. Hence, when we move from one spin pair to another in a regular lattice structure, their chirality $\vec S_j\times\vec S_{j'}$ minimizing the energy is constant in magnitude and antiparallel to $\vec u_{jj'}$. Therefore, each spin will be rotated with respect to its neighbors by a constant angle around a constant axis. The rotation direction is parallel or antiparallel to the spin-wave propagation direction $\vec q$, depending on whether $\vec u_{jj'}$ is parallel or antiparallel to $\tilde z$. In this way, a cycloidal structure with a period $\lambda$ forms in the spin system. Since the modulation occurs along the $[1\bar10]$ direction, this is where we observe the dispersion minimum shift by $q_\text{min} = 2 \pi/\lambda$ (Fig.~\ref{fig:figure2}). For the magnetization $\tilde z$ pointing along the $[1\bar10]$ direction, the $\vec u$ vectors (small arrows) cancel out when looking along this direction, and no frustrated structure of lower energy can arise. Since the Dzyaloshinskii-Moriya interaction operates in the sample plane, neither will we observe its hallmarks when $\tilde z$ is perpendicular to this plane.

\begin{figure}
\centering
  \includegraphics[width=\linewidth]{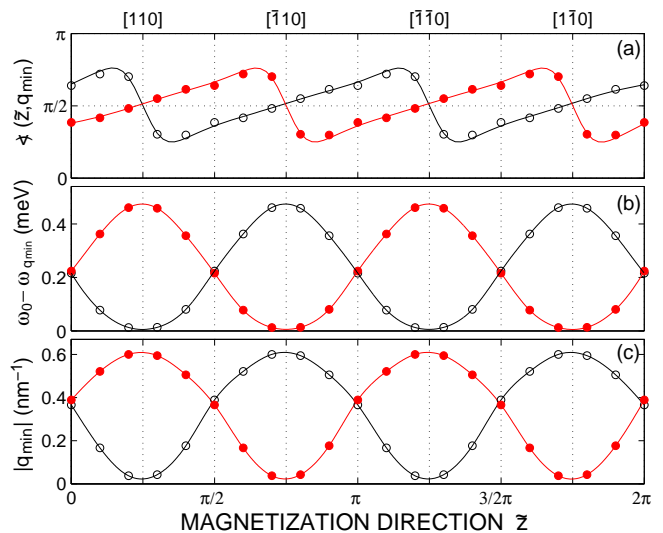}
\caption{(Color online) Parameters of the minimum shift $\vec q_\text{min}$ in two (Ga,Mn)As monolayers from Fig.~\ref{fig:figure2} as a function of the magnetization direction $\tilde z$ for the analyzed system (full circles) and for its mirror-reflection (open circles). a) The direction of the $\vec q_\text{min}$ vector with respect to $\tilde z$, b) energy gain $\omega_0 - \omega_\vec{q_\text{min}}$, c) magnitude of the minimum shift $\vec q_\text{min}$.}
\label{fig:figure6}
\end{figure}

For the quantitative analysis of the described effects, we fit the spin-wave dispersion presented in Fig.~\ref{fig:figure2} with the form~\eqref{eq:omegamacro}. In Fig.~\ref{fig:figure6} (full circles) we sweep the magnetization $\tilde z$ in the sample plane (001) and report the obtained angle between $\tilde z$ and $\vec q_\text{min}$ together with the energy gain $\omega_0 - \omega_\vec{q_\text{min}}$ due to the Dzyaloshinskii-Moriya interaction, and the magnitude of $\vec q_\text{min}$. From simple algebra we have $\vec q_\text{min} = -\frac{1}{2} \mathbf D^{-1} \mathbf U$ and $\omega_0 - \omega_\vec{q_\text{min}} = \vec q_\text{min}^\mathrm{T}\, \mathbf D\, \vec q_\text{min}$. As expected, the energy gain is the largest when $\tilde z$ and $\vec q_\text{min}$ are perpendicular to each other, for $\tilde z \parallel [110]$, and drops to zero for $\tilde z \parallel [1\bar10]$ (Fig.~\ref{fig:figure6}\,a,b). The first arrangement is accompanied by the strongest frustration of the spin system, $q_\text{min}  = 0.6$\,nm$^{-1}$ (Fig.~\ref{fig:figure6}\,c), resulting in the spin cycloid with period $\lambda = 10$\,nm. In the other arrangement, no spin frustration arises and the collinear spin order is preserved. This behavior accounts for an in-plane anisotropy with the easy and hard axes along the $[110]$ and $[1\bar10]$ directions, respectively. It is easy to notice that if we mirror-reflect the sample (or equivalently, move the bottom Ga1 layer to the top Ga5 position), the two axes switch---[110] will become the hard axis and $[1\bar10]$ the easy axis (Fig.~\ref{fig:figure6}, open circles). Consequently, the sign of $\vec q_\text{min}$ and the orientation of the spin cycloid will change. The described anisotropy arises from the net Dzyaloshinskii-Moriya interaction. Although it is a surface effect, it should not be confused with the surface anisotropy, which does not occur in the zinc-blende (001) surface.

In the above reasoning, it is important to remember that we deal with the hole-mediated ferromagnetism. While the magnetic lattice ions substitute only Ga sites and would seem oblivious to the inversion asymmetry of the Ga-As pairs, the carriers interact with all surrounding atoms. Thus, the system of magnetic ions interacting through the hole carriers, as described by Eq.~\eqref{eq:heisenberg}, is sensitive to all symmetry properties of the crystal.

As we have shown above, the Dzyaloshinskii-Moriya interaction in thin (Ga,Mn)As layers leads to the frustration of spins in the ground state. While in bulk (Ga,Mn)As the diagonal form of tensor $\mathbf A$~\eqref{eq:Abulk} depicts the uniform ground state magnetization, the exchange stiffness tensors in thin layers reveal a spin cycloid oriented along one of the mirror planes:
\begin{equation}
\begin{split}
&\mathbf A =
\begin{pmatrix}
0.93 & -0.28\\
-0.28 & 0.93
\end{pmatrix} \text{meV nm$^{-1}$}\,,\\
&\mathbf T =
\begin{pmatrix}
0.09-0.01i & 0.05+0.005i\\
0.05+0.005i & 0.09+0.006i
\end{pmatrix} \text{meV nm$^{-1}$}\,.
\end{split}
\end{equation}
Their eigenvectors point along the highest symmetry axes, [110] and $[1\bar10]$. This effect is equivalent to applying strain along one of these directions, which would account for a magnetoelastic anisotropy, like the one in biaxially strained samples. However, contrary to the case of a uniformly strained sample, all effects related to the Dzyaloshinskii-Moriya interaction vanish in the bulk limit.

The $\mathbf A$, $\mathbf T$ and $\vec U$ constants let us think of the magnetic system in terms of the interactions by which it is governed, as described by the Hamiltonian~\eqref{eq:Ha} in the interaction representation. Alternatively, if we want to deal with independent stationary magnons (with already deformed polarization), represented by the Hamiltonian~\eqref{eq:Hbb}, we calculate the spin-wave stiffness $\mathbf D$, related to $\mathbf A$ by the formula~\eqref{eq:coeff}. The difference between these two pictures is especially apparent at finite temperatures, where the length of the magnetization vector, $M(T)$, decreases and the formula becomes
\begin{equation}
	\label{eq:coeffT}
	\mathbf D(T) =\frac{2 g \mu_\text{B}}{M(T)} \mathbf A(T)\ .
\end{equation}
The higher the temperature, the stronger we have to tilt the magnetization vector in order to excite the quantum of spin waves (a magnon). Hence, $\mathbf A(T)$ decreases with temperature faster than $\mathbf D(T)$.

\subsection{Spin-wave stiffness and Curie temperature}
\label{sec:exchange}

This section discusses the relation between the spin-wave stiffness and Curie temperature, based on which we define the normalized spin-wave stiffness parameter, $\Dnor$. Depending only weakly on the hole density $p$ and spin splitting $\Delta$, it makes a convenient tool for experimentalists to estimate the spin-wave stiffness values $D$ given the Curie temperature and the magnetization of the sample. We provide quantitative numerical results on $\Dnor$ for bulk (Ga,Mn)As obtained in the $spds^\ast$ tight-binding computational scheme (see Sec.\,I), and compare them to the outcome of the previously employed six-band $\kp$ model.\cite{Konig:2001_PRB} We also clarify that the surprisingly large magnitude of $D$ in these systems results from the $p$-like character of the periodic part of Bloch function.

The standard relation between spin-wave stiffness $D$ and Curie temperature $T_\text{C}$ in a cubic crystal, which is often used in the literature on the topic, reads
\begin{equation}
\label{eq:D_Tc1}
D = \frac{k_\text{B}T_\text{C} r_\text{nn}^2}{2(S+1)} \ ,
\end{equation}
where $r_\text{nn}$ is the nearest neighbor distance.\cite{Skomski:2008_B} Derived for short-range interactions, it is interesting to find out how it is modified when considering the actual nature of the spin-spin exchange. This question has a number of implications. For instance, both carrier relaxation time, which is limited by magnon scattering, and the quantitative accuracy of the mean-field approximation grow with $D/T_\text{C}$, as the density of spin waves at given temperature $T$ diminishes when $D$ increases. It is worth noting that a simple adaption of this formula to (Ga,Mn)As by replacing $r_\text{nn}$ with $(xn_0)^{-1/3}$ resulted in an overestimation of the $\pd$ exchange integral by an order of magnitude.\cite{Wang:2007_PRB}

Allowing for spatially modulated structures, the magnetic ordering temperature $T_\text{C}$ for a carrier-controlled ferromagnet is given by a solution of the mean-field equation,\cite{Matsukura:2002_B, Dietl:1999_MSE}
\begin{equation}
\label{eq:mfeq}
\beta^2 \chi(\vec q,T) \chi_S(\vec q,T) = 1 \ ,
\end{equation}
where $\chi$ and  $\chi_S$ are the carrier and lattice ion spin susceptibilities. In the simple case of a parabolic band they are proportional to the Pauli and Curie-Weiss magnetic susceptibilities, respectively.

First, we consider the two-band model of carriers residing in a simple parabolic band, where all that is left of the $\boldsymbol{\mathcal A}$ tensor of Eq.~\eqref{eq:EgeneralAK} is the scalar isotropic exchange stiffness $A$. According to the previous section, it is related to the quadratic coefficient of the expansion $\chi(q) \approx \chi_0 + n S \beta^2 C q^2$, by $D = n S \beta^2 C = 2A/nS$. Additionally, we assume that the ground state of the system corresponds to the uniform ferromagnetic ordering, $\vec q=0$, and take $\chi_S(\vec q,T)$ in the Curie form,
\begin{equation}
\label{eq:chiS}
\chi_S(\vec{q},T) = \frac{nS(S+1)}{3k_\text{B}T} \ .
\end{equation}
Using Eqs.~\eqref{eq:mfeq} and \eqref{eq:chiS}, we obtain
\begin{equation}
D = \frac{3k_\text{B}T_\text{C}\,C(T\to0)}{(S+1)\,\chi(0,T=T_\text{C})} \ .
\end{equation}
If the values of both spin splitting $\Delta$ at $T \to0$ and $k_\text{B}T_\text{C}$ are much smaller than the magnitude of the Fermi energy $|E_\text{F}|$, the carrier susceptibility is given by
\begin{equation}
\label{eq:chi2b}
\chi(q) = \frac{1}{4}\rho(E_\text{F})\ F\left(\frac{q}{2k_\text{F}}\right) \ .
\end{equation}
The total density of states at $E_\text{F} = \hbar^2 k_\text{F}^2/(2 m^\ast)$ is $\rho(E_\text{F}) = m^\ast k_\text{F}/(\pi\hbar)^2$, where $m^\ast$ is the carrier effective mass, and
\begin{equation}
\label{eq:lindhard}
\begin{split}
F\left(\frac{q}{2k_\text{F}}\right)
&=\frac{1}{2} + \frac{k_\text{F}}{2q}\left(1- \frac{q^2}{4k_\text{F}^2}\right)\log\left|\frac{2k_\text{F}+ q}{2k_\text{F}-q}\right| \\
&= 1 - \frac{q^2}{12k_\text{F}^2} - O\left(\frac{q^4}{k_\text{F}^4}\right)
\end{split}
\end{equation}
is the Lindhard function. We obtain from these equations
\begin{equation}
\label{eq:D_Tc2}
D = \frac{k_\text{B}T_\text{C}}{4(S+1)k_\text{F}^2} \ .
\end{equation}
We see that, in agreement with the notion that magnetic stiffness increases with the range of the spin-spin interaction, $r_\text{nn}$ in Eq.~\eqref{eq:D_Tc1} is replaced by $1/(k_\text{F}\sqrt{2})$ in Eq.~\eqref{eq:D_Tc2}, which scales with the range of the carrier-mediated interactions. Indeed, according to the Ruderman-Kittel-Kasuya-Yosida theory,\cite{Ruderman:1954_PR} the magnitude of the ferromagnetic exchange integral decays at small spin-spin distances $r$ as $1/(rk_\text{F})$ and reaches the first zero at $r \approx 2.2/k_\text{F}$.

Since in semiconductors $1/k_\text{F} \gg a_0$, the above formulae imply that $D/T_\text{C}$ is rather large in systems with carrier-controlled ferromagnetism. A question arises as to how the ratio $D/T_{\text{C}}$ would be modified, if the complexities of the valence band were taken into account.

As already noticed by K\"onig {\em et al.},\cite{Konig:2001_PRB} the values of exchange stiffness for (Ga,Mn)As are greater in the six-band model with a spin-orbit coupling than in the case of a simple parabolic band. As argued by these authors, due to the multiplicity of the valence bands, the Fermi level for a given carrier concentration is much lower than in the two-band model, and hence both carrier polarization and exchange stiffness are greater.\cite{Konig:2001_PRB} On the other hand, Brey and G\'omez-Santos\cite{Brey:2003_PRB} assign large values of $D$ to higher magnitudes of $T_\text{C}$ in the multi-band model.

\begin{figure}
\centering
  \includegraphics[width=0.95\linewidth]{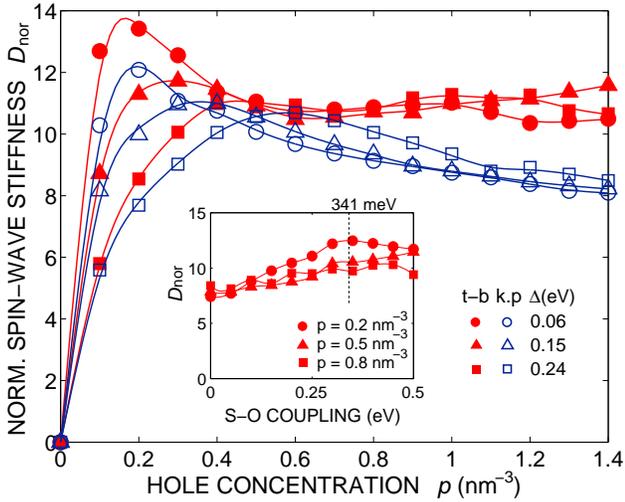}
\caption{(Color online) Normalized spin-wave stiffness $\Dnor$ in bulk (Ga,Mn)As as a function of the hole concentration $p$ and the spin-orbit coupling (inset; the value of the spin-orbit coupling in (Ga,Mn)As is indicated by dotted line) for different spin splittings $\Delta$. Filled and empty markers indicate the results of the $spds^\ast$ tight-binding and the six-band $\kp$ model, respectively.}
\label{fig:figure7}
\end{figure}

To check these suggestions we plot in Fig.~\ref{fig:figure7} the values of dimensionless parameter
\begin{equation}
\Dnor = \frac{4(S+1)k_\text{F}^2D}{k_\text{B}T_\text{C}}\ ,
\end{equation}
for various hole concentrations $p$ and $k_\text{F} = (3\pi^2p)^{1/3}$. The ratio $D/T_\text{C}$ for bulk (Ga,Mn)As is computed using the $spds^\ast$ tight-binding (Sec.~\ref{sec:section1}) and the previously employed\cite{Konig:2001_PRB} six-band $\kp$ model of the semiconductor band structure. Guided by the results of the two band model, one could expect $\Dnor \lesssim 1$ in the limit of small spin polarizations, $\Delta \ll |E_\text{F}|$. In contrast to these expectations, $\Dnor$ is of the order of eleven and, moreover, weakly dependent on the polarization, altered here by changing the hole concentration and the spin-orbit splitting. Furthermore, the experimentally observed biaxial strain magnitudes have only slight effect on the stiffness tensor (see Fig.~\ref{fig:figure1} and Eq.~\eqref{eq:Abulk}.\cite{Konig:2001_PRB} This indicates that the single parameter $\Dnor$ can serve to estimate the magnitudes of $D$ and $A$ if only the Mn magnetization $M$ and hole concentration $p$ are known.

\begin{figure}
\centering
  \includegraphics[width=0.99\linewidth]{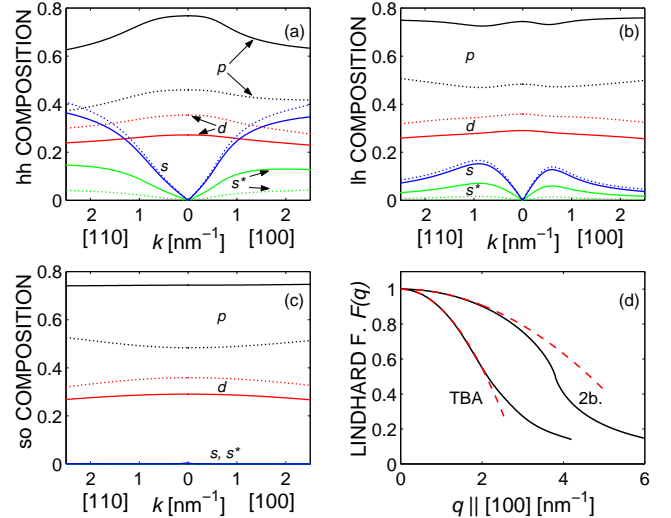}
  \caption{(Color online) Atomic composition of heavy, light and spin-orbit split-off hole bands (a, b and c panels) as a function of the $k$-vector. It is given by the squares of norms of projections of the carrier state on atomic $s$, $p$, $d$ or $s^\ast$ orbitals of As or Ga atoms (solid and dotted lines). Panel d presents the Lindhard function $F(q)$ calculated in the tight-binding model ($p=0.65$\,nm$^{-3}$ and $\Delta = -0.13$\,eV, with the spin-orbit coupling set to zero) and in the parabolic band model with the same Fermi energy $E_\text{F} = -0.27$\,eV. Dashed lines are their square form fits.}
\label{fig:figure8}
\end{figure}

Knowing that neither the spin-orbit coupling nor the multiplicity of carrier bands can explain the large spin-wave stiffness, we turn our attention to the matrix elements $\langle u_{\vec{k},m}|\hat{s}^\sigma|u_{\vec{k+q},m'}\rangle$ in the spin susceptibility of holes~\eqref{eq:chi}. Neglecting the spin-orbit coupling, $u_{\vec{k},m}$ is a product of the spin part $s_m$ and the real spatial part $w_{\vec{k},m}$. Thus, we can write
\[
\langle u_{\vec{k},m}|\hat{s}^\sigma|u_{\vec{k+q},m'}\rangle = \pm\frac{1}{\sqrt{2}} (1-\langle s_m|s_{m'}\rangle) \langle w_{\vec{k},m}|w_{\vec{k+q},m'}\rangle
\]
for $\sigma = +,-$. In the parabolic two-band model with its carrier wave functions described by plane waves $\psi_{\vec{k},m}(\vec r) = \frac{s_m}{\sqrt{V}}\, e^{i\vec k\cdot\vec r}$, the periodic part $u_{\vec{k},m} = \frac{s_m}{\sqrt{V}}$. Hence $\langle s_m|s_{m'}\rangle=\delta_{mm'}$ and $w_{\vec{k},m}=\frac{1}{\sqrt{V}}$, and $F(\vec q) = \chi({\vec{q}})/\chi(0)$ simplifies to the Lindhard function~\eqref{eq:lindhard}. More realistic models take into account the periodic lattice potential, which mixes different atomic orbitals and leads to the $\vec k$-dependent modulation of $u_{\vec k,m}$. (The eventual composition of hole states in the $spds^\ast$ tight-binding model is presented in Fig.~\ref{fig:figure8}\,a-c.) From this it follows that the scalar products $\langle w_{\vec{k},m}|w_{\vec{k+q},m'}\rangle$ are $\vec q$-dependent, and can be only smaller than in the simple model. As a consequence, $F(\vec q)$ has a steeper slope, as presented in Fig.~\ref{fig:figure8}\,d. This explains why the magnitude of the stiffness tensor is so large in (Ga,Mn)As and related ferromagnets. Actually, the fact that the $p$-type character of the carrier wave functions affects in this way the $q$-dependence of dielectric susceptibility has been already noted.\cite{Broerman:1971_PRB, Szymanska:1978_JPCS}

\subsection{Spin waves' contribution to magnetization}
\label{subsec:magntemp}

In this section, we address the problem of applicability of the mean-field model to the description of temperature dependence of magnetization in the analyzed systems. While it is a known fact that the critical fluctuations do not change much the mean-field Curie temperature $T_C$ in the presence of long-range exchange interactions,\cite{Binney:1992_B} several papers\cite{Konig:2000_PRL, Callen:1963_PR, Yang:2001_PRL} discuss how it is lowered by spin waves. Also, the influence of dilution and disorder on spin waves and the Curie temperature, which can be crucial in samples with low Mn content, was studied by many authors with varying conclusions.\cite{Berciu:2002_PRB, Meilikhov:2007_PRB, Singh:2007_JP, Szalowski:2008_PRB, Tang:2008_PRB, Balcerzak:2009_PRB} In response, we propose a simple model which takes into account the correct number of spin-wave excitations and thermal depolarization of spins. However, we argue that spin waves also do not lower the mean-field $T_C$ due to thermal decoherence of quantum spin system.

The mean-field approximation employed in our model allows for a reasonable overall description of ferromagnetism in the analyzed systems. It reduces the problem of lattice spins coupled by the exchange interaction to that of noninteracting spins in the molecular field $\Xi$~\eqref{eq:Xi}. Their magnetization is described by the self-consistent equation,
\begin{equation}
\label{eq:BSmf}
M(T) = M(0)\, \mbox{B}_S\left( \frac{S\,\Xi(\Delta(T),T)}{k_\text{B} T} \right) \ .
\end{equation}
The Brillouin function $\mbox{B}_S$ ignores the actual nature of thermal fluctuations and their correlations, and assumes that every spin fluctuates independently, which in our theory corresponds to the high-$q$ limit of spin-wave excitations (see Fig.~\ref{fig:figure1}). At low temperatures, however, the long-wavelength magnons of much lower energies can exist. Since each of them carries a moment of $\mu_\text{B}$, they yield a strong contribution to the temperature dependence of magnetization,\cite{Skomski:2008_B, Bloch:1930_ZP}
\begin{equation}
\label{eq:MTsw}
M(T) = M(0) \Bigl( 1 - (NS)^{-1} \sum_{\vec q}\langle n_{\vec q}\rangle_T\Bigr) \ ,
\end{equation}
where $\langle n_{\vec q}\rangle_T$ is the thermal average of spin-wave excitations in each mode. The latter can be modeled by the Bose-Einstein distribution, $\langle n_{\vec q}\rangle_T = 1/(\exp(\omega_{\vec q}/k_\text{B}T) - 1)$, as spin waves are bosons. Then, replacing the summation in Eq.~\eqref{eq:MTsw} by an integral and putting $\omega_{\vec q} \approx D q^2$, one obtains the well known $T^{3/2}$ Bloch law\cite{Kittel:1987_B}
\begin{equation}
\label{eq:nq1}
\int \der \vec q\ \langle n_{\vec q}\rangle_T = \zeta_{3/2}\,\pi^{3/2} \left(\frac{k_\text{B}T}{D}\right)^{3/2},
\end{equation}
where $\zeta_{3/2} \approx 2.612$ is the Riemann zeta function.

Remembering about the spin-wave gap created by magnetic anisotropy, which allows for ferromagnetism in low dimensional systems despite the Mermin-Wagner theorem,\cite{Mermin:1966_PRL} we have $\omega_{\vec q} \approx \omega_0 + D q^2$. (We neglect the Dzyaloshinskii-Moriya coefficient, which vanishes in bulk limit.) In that case, the above law is modified to\cite{Jackson:1989_JMMM}
\begin{equation}
\label{eq:nq2}
\int \der \vec q\ \langle n_{\vec q}\rangle_T = \mathrm{Li}_{3/2}\left(e^{-\omega_0/k_\text{B}T}\right)\pi^{3/2} \left(\frac{k_\text{B}T}{D}\right)^{3/2},
\end{equation}
where $\mathrm{Li}_{3/2}\left(e^{-\omega_0/k_\text{B}T}\right)$ is de Jonqui\'ere's function.

Furthermore, the Bose-Einstein statistics allows for the unlimited number of spin waves in each mode (on the other hand, K\"onig \etal\cite{Konig:2001_PRB}\ assume that there can be only $2S$ spin waves of each $\vec{q}$, which is too strict). In fact, their total number cannot be larger than $2NS$, corresponding to complete magnetization reversal. We handle this by introducing a fictious mode with zero energy, which is `occupied' by the spin waves which have not been excited in reality. In this way we can treat the problem with classic Bose-Einstein condensate methods:\cite{Huang:1987_B} the total number of bosons occupying all modes is always $2NS$, their zero-energy mode constitutes the `condensate' phase, while the excited spin waves constitute the `thermal cloud'. The total number of spin waves in the limit of infinite crystal volume is therefore given by
\[
\min \left( \sum_{\vec{q} \le \vec q_\text{D}} \frac{ e^{-\omega_\vec{q}/k_\text{B}T} }{1 - e^{-\omega_\vec{q}/k_\text{B}T}} , 2 N S \right) \ .
\]

Additionally, while at zero temperature the system is described by a pure state $\Psi_0$, where exciting a spin wave costs the energy
\begin{equation}
	\omega_{\vec q}^0 = \langle b^\dagger_{\vec q} \Psi_0 | \mathcal H^\text{eff} | b^\dagger_{\vec q} \Psi_0 \rangle - \langle \Psi_0 | \mathcal H^\text{eff} | \Psi_0 \rangle\ ,
\end{equation}
at non-zero temperatures it is described by a mixture of pure states $\Psi_n$ with a certain number of spins flipped by one-particle thermal excitations, $\rho = \sum_n p_n |\Psi_n \rangle \langle \Psi_n|$. Thus, its magnetization drops to $M(T)$ according to the mean-field Brillouin function~\eqref{eq:BSmf}. The energy cost of exciting a spin wave is now given by
\begin{equation}
	\omega_{\vec q}^T = \sum_n p_n \left[\langle b^\dagger_{\vec q} \Psi_n | \mathcal H^\text{eff} | b^\dagger_{\vec q} \Psi_n \rangle - \langle \Psi_n | \mathcal H^\text{eff} | \Psi_n \rangle\right]\ .
\end{equation}
Since the spin-wave dispersion $\omega_{\vec q}$~\eqref{eq:omega} depends on temperature $T$ almost exclusively via the spin splitting $\Delta$ and is approximately proportional to it, we can estimate the above expression by $\omega_{\vec q}^0 \sum_n p_n \frac{M_n}{M(0)} = \omega_{\vec q}^0 \frac{M(T)}{M(0)} = \omega_{\vec q}^0 \frac{\Delta(T)}{\Delta(0)}$, where $M_n$ is the magnetization in the state $\Psi_n$. Spin waves are thus perturbations of the thermal state of lattice spins and not of the ground state, which we can model just by replacing $M(0)$ with $M(T)$ (or $\Delta$ with $\Delta(T)$) in the dispersion relation $\omega_\vec{q}$. They additionally lower the magnetization to $M'(T)$. Remaining in the limit of small oscillation approximation, we obtain the following set of equations:
\begin{equation}
\begin{split}
\label{eq:swbec}
M(T) &= M(0) \mbox{B}_S \left( \frac{S \Xi(\Delta'(T))}{k_\text{B} T} \right) \ , \\
M'(T) &= M(T) \\
&\ - \min \left( \frac{M(0)}{N S} \sum_{\vec{q} \le \vec q_\text{D}} \frac{ e^{\xi_\vec q} }{1 - e^{\xi_\vec q}}, 2 M(T) \right) \,,
\end{split}
\end{equation}
where $\xi_\vec q = -\omega_\vec{q}(\Delta(T), T)/k_\text{B}T$. The spin splitting $\Delta(T)$ is induced by the mean field of the lattice spins described by the first equation, while $\Delta'(T)$ additionally takes into account their depolarization due to spin waves.
\begin{figure}
\centering
  \includegraphics[width=0.9\linewidth]{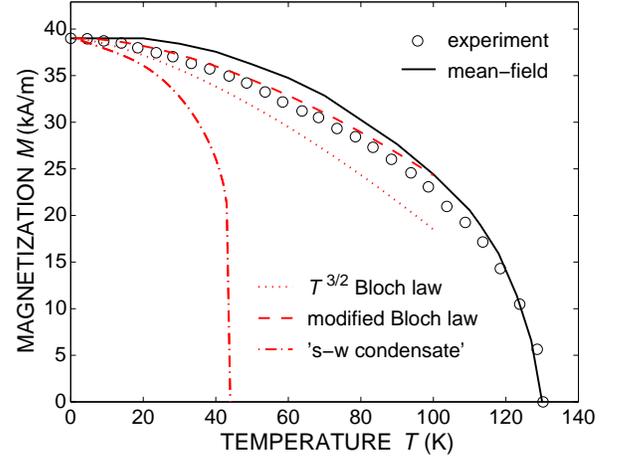}
  \caption{(Color online) Temperature dependence of magnetization for bulk (Ga,Mn)As from Fig.~\ref{fig:figure1}, calculated by the mean-field model~\eqref{eq:BSmf}, $T^{3/2}$ Bloch law~\eqref{eq:MTsw} and its modified version~\eqref{eq:nq2}, and the `spin-wave condensate'~\eqref{eq:swbec}. The experimental dependence for the sample from Ref.~\onlinecite{Gourdon:2007_PRB} is analyzed in Sec.~\ref{sec:experimental}.}
\label{fig:figure9}
\end{figure}

Figure \ref{fig:figure9} presents the temperature dependence of magnetization as described by the above methods for the bulk (Ga,Mn)As from Fig.~\ref{fig:figure1}. The $T^{3/2}$ Bloch law and its modified version apply to low temperatures only. We see that the values of $D$ determined neglecting the anisotropy gap would be overestimated. The `spin-wave condensate' includes the magnon contribution to magnetization in the whole temperature range, and leads to significantly lower Curie temperature than in the mean-field model. This method uses the correct bound for the number of spin-wave excitations. It also naively attempts to solve the problem of the well-known shortcomings of the spin-wave theory\cite{Aharoni:2001_B} introduced by the Holstein-Primakoff transformation~\eqref{eq:spinoperators}, as it includes thermal disorder by depolarizing the lattice spins with temperature. However, quantum-mechanical intuition suggests that the spin waves should vanish completely at higher temperatures. Non-zero temperature leads to the loss of information about the system: pure states are replaced by mixed states, namely the thermal states $e^{-\beta H} / \text{Tr}\, e^{-\beta H}$, and the quantum correlation of spins vanishes. This can be reflected in the $\pd$ Hamiltonian by replacing $\beta \sum_{i,j} \vec s_i \cdot \vec S_j$ with $\beta \sum_{i,j} \langle s^z_i \rangle \langle S^z_j \rangle = \Xi \sum_{j} \langle S^z_j \rangle$, which is simply the mean-field model. Similarly, in the vicinity of the Curie transition, where the problem of critical fluctuations arises, the mean-field models of systems with long-range exchange interactions are known to work very well.\cite{Binney:1992_B}

The next section provides further insight into the applicability of the presented models by comparing their results to experiment.

\section{Comparison to experimental data}
\label{sec:experimental}

As already mentioned in Sec.~I, the theory of exchange stiffness developed within the six-band $\pd$ Zener model\cite{Konig:2001_PRB} describes quantitatively the width of stripe domains in (Ga,Mn)As.\cite{Dietl:2001_PRBb} Recently, Gourdon \etal\cite{Gourdon:2007_PRB} carried out a detailed analysis of the magnetic domain structure and magnetic properties of an annealed 50\,nm-thick Ga$_{0.93}$Mn$_{0.07}$As layer with a perpendicular magnetic easy axis and the Curie temperature of 130\,K. Two employed experimental methods yielded an upper and lower limit of the isotropic exchange stiffness $A$ as a function of temperature $T$. As found by these authors from examining the domain-wall velocity, the higher values of $A(T)$, determined from the lamellar domain width, are reliable.

We model the sample in the tight-binding computational scheme for bulk (Ga,Mn)As (see Sec.~\ref{sec:section1}). In order to determine the material parameters for numerical simulations, we start by estimating the effective Mn content $x_\text{eff}$ from the measured low-temperature spontaneous magnetization $M(T\to0) = 39$\,kA/m. Taking into account the hole contribution, $M_c \approx -5$\,kA/m,\cite{Sliwa:2006_PRB} implies the magnetization of the Mn spins $M_S = 44$\,kA/m. This value corresponds to the effective Mn content $x_\text{eff} = 4.3\%$ and the spin splitting $\Delta = -0.13$\,eV. No direct measurements of the hole concentration are available for this sample, so we estimate its magnitude from the effective and total Mn content, $x_\text{eff}$ and $x = 7\%$. Assuming that interstitial Mn donors had survived the annealing process and formed antiferromagnetic pairs with the substitutional Mn acceptors,\cite{Blinowski:2002_PRB, Edmonds:2004_PRL} we obtain $p = (3x_\text{eff}/2-x/2) n_0 = 0.65$\,nm$^{-3}$.\cite{Stefanowicz:2010_PRB}

From Fig.~\ref{fig:figure7}, for the given value of $p$ and $\Delta$ we find $\Dnor \approx 10.5$, which gives the spin-wave stiffness $D = 1.1$\,meV\,nm$^2$ ($A = 0.21$\,pJ/m) at $T \to 0$\,K. Knowing this, we can calculate the temperature dependence of magnetization according to Sec.~\ref{subsec:magntemp} (Fig.~\ref{fig:figure9}). In the mean-field picture (solid lines), the magnetization of lattice spins $M_S$ is described by the Brillouin function~\eqref{eq:BSmf}. The magnitude of hole magnetization $|M_c|$ decreases with temperature proportionally to $M_S$, according to Eq.~\eqref{eq:Xi}. The resulting magnetization $M(T)$ is compared to the experimental curve (circles). We obtain a good agreement with the measured data, especially near the Curie transition, and $T_\text{C} = 127$\,K. At low temperatures, we plot the outcomes of the $T^{3/2}$ Bloch law~\eqref{eq:nq1} and its modified version~\eqref{eq:nq2} employing the calculated spin-wave stiffness value $D$ (dotted lines). The modified Bloch law, adjusted to include the spin-wave gap, gives very good agreement with the experimental dependence, which indicates that in this regime the spin-wave excitations are solely responsible for demagnetization. Near the Curie transition, we reconstruct the measured Curie temperature and magnetization values with the Brillouin function, which suggests that the temperature destroys the spin-wave coherence and recalls the mean-field picture.

\begin{figure}
\centering
  \includegraphics[width=0.9\linewidth]{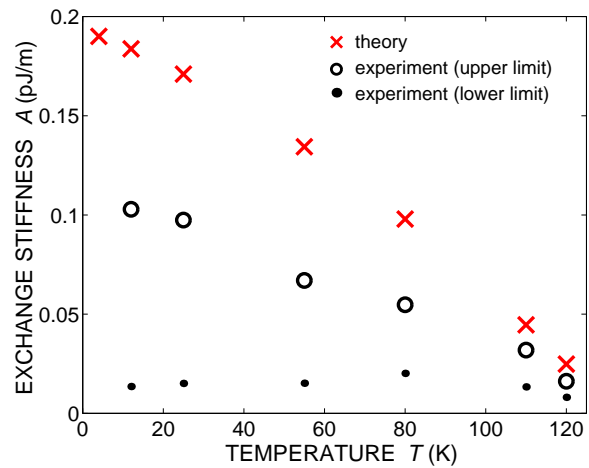}
\caption{(Color online) Theoretical reconstruction of experimental data on exchange stiffness $A$ as a function of temperature $T$ obtained in Ref.~\onlinecite{Gourdon:2007_PRB} from the analysis of the lamellar domain width in Ga$_{0.93}$Mn$_{0.07}$As.}
\label{fig:figure10}
\end{figure}

To reconstruct the $A(T)$ trend obtained from the magnitudes of lamellar domain width,\cite{Gourdon:2007_PRB} we again make use of the fact that the susceptibility $\chi^{+-}_{\vec q}$~\eqref{eq:chi} depends on temperature almost exclusively via spin splitting. Thus, and according to the formula~\eqref{eq:coeffT}, the exchange stiffness scales with temperature as $\Delta(T)^2$.\cite{Dietl:2001_PRBb} We use the experimentally determined $M(T)$ and the calculated $A(T\to0)=0.21$\,pJ/m to estimate the exchange stiffness values for the remaining temperatures as $A(T) = A(0) M(T)/M(0)$.
As shown in Fig.~\ref{fig:figure10}, this procedure correctly reproduces the experimental $A(T)$ trend. However, the same value of $A(T\to0)$, which has been shown to successfully describe the experimental $M(T)$ dependence with the modified Bloch law, is twice as large as the exchange stiffness constant determined from the measured lamellar domain width.

Potashnik \etal\cite{Potashnik:2002_PRB} evaluated the isotropic exchange constant $J$ for the set of optimally annealed (Ga,Mn)As layers with varying Mn content. The values derived from the temperature dependence of magnetization, using the standard Bloch law, and from the Curie temperature within the three-dimensional Heisenberg model,\cite{Kittel:1987_B} were similar.

\begin{figure}
\centering
  \includegraphics[width=0.9\linewidth]{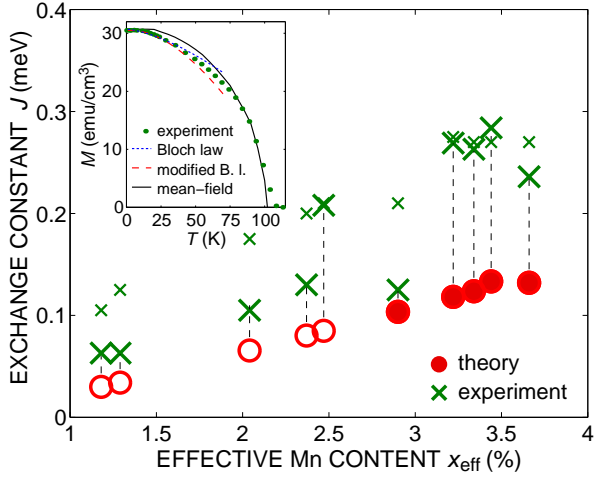}
\caption{(Color online) Theoretical reconstruction of experimental data on exchange constant $J$ for samples with varying effective Mn content $x_\text{eff}$, as determined from the temperature dependence of magnetization using the $T^{3/2}$ Bloch law (large crosses) and from the Curie temperature (small crosses) in Ref.~\onlinecite{Potashnik:2002_PRB}. Empty circles indicate the region where the experimental $T_C$ exceeds the theoretical predictions. Inset: experimental temperature dependence of magnetization for the sample with $x_\text{eff} = 3.3\%$ fit by the standard Bloch law with $J = 0.26$~meV (Ref.~\onlinecite{Potashnik:2002_PRB}), the modified Bloch law~\eqref{eq:nq2} with the theoretically obtained $J = 0.12$~meV and by the mean-field Brillouin function.}
\label{fig:figure11}
\end{figure}

Our theoretical reconstruction of the experiment by the $spds^\ast$ tight-binding model of bulk (Ga,Mn)As is demonstrated in Fig.~\ref{fig:figure11}. To obtain the presented results, we have estimated the effective Mn content $x_\text{eff}$ from the measured low-temperature magnetization $M(T\to0)$, and then increased it by about $10\%$ to include the hole contribution.\cite{Sliwa:2006_PRB} Assuming that annealing removed all interstitials and each remaining substitutional Mn produces one hole carrier, the hole density $p = x_\text{eff}\, n_0$. For the obtained values of $p$ and $x_\text{eff}$ we have calculated the mean-field Curie temperatures $T_\text{C}$ and the exchange constant $J = D (x_\text{eff}\,n_0)^{2/3}/2 S$ assuming that Mn ions form a cubic lattice. We note that this form is equivalent up to a few percent to that employed in Ref.~\onlinecite{Potashnik:2002_PRB}: $J = D(4\pi x n_0/24)^{2/3}/2S$, where $x$ is the total Mn content (N.~Samarth, private communication). Since for the samples with low Mn content our values of $T_\text{C}$ are much lower than the experimental ones, we conclude that the corresponding estimates of $J$ (indicated by empty circles) are not reliable. On the other hand, we reconstruct Curie temperatures for samples with $x_\text{eff} > 2.5\%$ (filled circles), but this time the theoretical values of $J$ are much smaller than the experimental ones. This discrepancy points to the importance of the anisotropy-induced energy gap in the spin-wave spectrum. As illustrated in Fig.~\ref{fig:figure11} (inset) for the sample with $x_\text{eff} = 3.3\%$, fitting the experimental $M(T)$ curves with the standard Bloch law neglecting the gap leads to higher values of $J$ than those expected theoretically. At the same time, the modified Bloch law~\eqref{eq:nq2} employing the theoretical exchange constant $J = 0.12$~meV\,nm$^2$ reconstructs the analyzed $M(T)$ trend. We notice that it perfectly describes the mild slope of the low-temperature $M(T)$, contrary to the standard Bloch law.\cite{Potashnik:2002_PRB} Similarly to the case of the Gourdon \etal\ experiment,\cite{Gourdon:2007_PRB} the mean-field model works very well at higher temperatures.

\begin{figure}
\centering
\includegraphics[width=\linewidth]{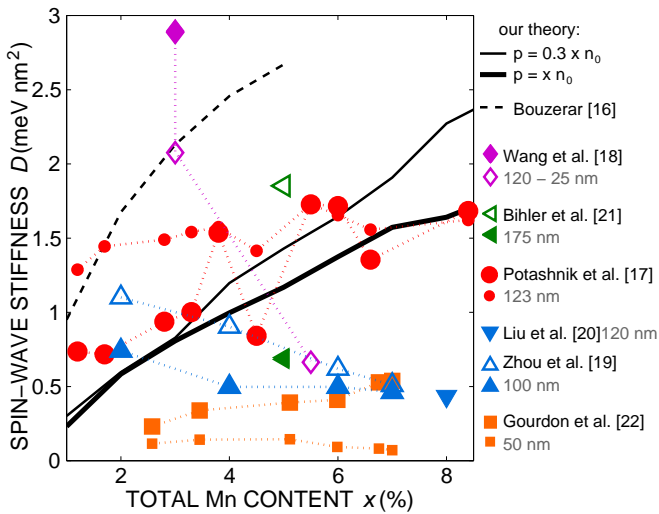}
\caption{(Color online) Compilation of theoretical results (lines), obtained with no adjustable parameters, and experimental data (markers) on the spin-wave stiffness $D$ as a function of the nominal Mn concentration $x$ in Ga$_{1-x}$Mn$_{x}$As. Solid symbols denote annealed as well as etched and hydrogenated samples (Ref.~\onlinecite{Bihler:2009_PRB}). Empty symbols denote as-grown samples. The value $g = 2$ was used to convert $D$ in magnetic units (T\,nm$^2$) to energy units (meV\,nm$^2$).}
\label{fig:figure12}
\end{figure}

In a series of experiments, the spin-wave stiffness was determined by examining spin precession modes excited by optical pulses\cite{Wang:2007_PRB} and under ferromagnetic resonance conditions.\cite{Zhou:2007_IEEE, Liu:2007_PRB, Bihler:2009_PRB} According to these works, the experimental findings are strongly affected by gradients of magnetic anisotropy, presumably associated with carrier depletion at the surface and interface, which also affect the character of spin pinning. We also note that no influence of the magnetic field on the hole spins, visible as a deviation of the Land\'e factor from the value $g = 2$,\cite{Sliwa:2006_PRB} was taken into account in the employed Landau-Lifshitz equations. With these reservations we show in Fig.~\ref{fig:figure12} the experimentally evaluated values of $D$ plotted as a function of the nominal Mn concentration $x$ for various as-grown and annealed samples of Ga$_{1-x}$Mn$_x$As. These findings are compared to the results of {\em ab initio} computations\cite{Bouzerar:2007_EPL} (dashed line) and our theory for the hole concentration $p = xn_0$ and $p=0.3xn_0$ (solid lines). When comparing theoretical and experimental results, one should take into account that the actual Mn concentration $x_\text{eff}$ is smaller than $x$, particularly in as-grown samples. As seen, our theory describes properly the order of magnitude of the spin-wave stiffness $D$ but cannot account for a rather large dispersion in the experimental data.

In thin (Ga,Mn)As layers, described in Sec.~\ref{sec:layers}, the Dzyaloshinskii-Moriya interaction can be observed in form of a spin-wave dispersion minimum shift. It leads to the formation of a cycloidal spin structure and uniaxial in-plane anisotropy of the [110] and $[1\bar10]$ directions, with the easy axis determined by the sample geometry. Both these phenomena have been a subject of considerable interest in recent years.\cite{Rappoport:2004_PRB, Zarand:2002_PRL, Sawicki:2004_PRB, Welp:2004_APL, Sawicki:2005_PRB} While the long-period spin structures have not been hitherto observed, the uniaxial anisotropy is commonly present in (Ga,Mn)As.\cite{Zemen:2009_PRB, Stefanowicz:2010_PRB} However, it was shown by gradual etching\cite{Sawicki:2005_PRB} or by investigating different samples\cite{Welp:2004_APL} to be insensitive to the layer thickness. This is not the case for the anisotropy caused by the Dzyaloshinskii-Moriya interaction, which is a surface effect and vanishes with increasing layer thickness. To test our theory, one should pick very thin high-quality samples grown in the [001] direction, which additionally excludes the anisotropy of the surface, with the easy axis along one of the in-plane diagonal directions. The cycloidal spin structure could be then observed e.g. under a magnetic force microscope or by neutron scattering. The sample should be probed along the [110] and $[1\bar10]$ directions to find the long-period modulation of magnetization only along the one perpendicular to the easy axis.

\section{Summary} 

We have investigated spin waves and exchange stiffness in thin layers and bulk crystals of ferromagnetic (Ga,Mn)As described by the $spds^\ast$ tight-binding computational scheme. Using the proposed variational-perturbation calculus, we have described the analyzed systems and their spin-wave excitations. Their properties have been expressed by the phenomenological parameters of micromagnetic theory. We have noticed that the strength of ferromagnetic order described by the isotropic exchange stiffness  is significantly amplified by the $p$-like character of carrier wave functions, as compared to the simple parabolic band model. Furthermore, we have found various effects reflecting the tendency of the spin-orbit interaction to pervade every aspect of carrier dynamics. They produce the relativistic corrections to spin-waves given by the anisotropic exchange stiffness tensor and the asymmetric Dzyaloshinskii-Moriya coefficient. The latter accounts for the cycloidal spin arrangement and the accompanying uniaxial in-plane anisotropy of diagonal ([110]/$[1\bar10]$) directions in thin layers, which can result in a surface-like anisotropy in thicker films. Quantitative results on the stiffness constant have been provided in form of a normalized parameter, which assumes the value $D_\text{nor} \approx 11$ over a wide range of Mn and hole concentrations in (Ga,Mn)As. They agree with the previous $\kp$ calculations\cite{Konig:2001_PRB, Brey:2003_PRB} but predict significantly smaller values of spin-wave stiffness than those resulting from {\em ab initio} computations.\cite{Bouzerar:2007_EPL} 

The above theories have been applied to analyze the related experimental data on the stiffness parameter and the temperature dependence of magnetization. Our basic theoretical model has not managed to reconstruct all stiffness values obtained by various experimental methods. In all cases we have reconstructed the entire range of magnetization dependence on temperature. At low temperatures, it can be understood within the the modified Bloch law\cite{Jackson:1989_JMMM} employing the values of spin-wave stiffness calculated by our model. At higher temperatures, the mean-field theory becomes justifiable owing to thermal decoherence and the long range character of spin-spin interactions. At the same time, we have reproduced only partly the stiffness values obtained from analyzing precession modes in (Ga,Mn)As thin films. Our results may allow to separate bulk and surface effects, as well as bring to light the pining phenomena and the role of inhomogeneities in experiments examining precession modes in slabs of carrier-controlled ferromagnetic semiconductors.

\begin{acknowledgments}
A.\,W. acknowledges support by the President of Polish Academy of Sciences and EC Network SemiSpinNet (PITN-GA-2008-215368), and T.\,D. acknowledges support from the European Research Council within the ``Ideas'' 7th Framework Programme of the EC (FunDMS Advanced Grant).
\end{acknowledgments}


\begin{thebibliography}{72}
\expandafter\ifx\csname natexlab\endcsname\relax\def\natexlab#1{#1}\fi
\expandafter\ifx\csname bibnamefont\endcsname\relax
  \def\bibnamefont#1{#1}\fi
\expandafter\ifx\csname bibfnamefont\endcsname\relax
  \def\bibfnamefont#1{#1}\fi
\expandafter\ifx\csname citenamefont\endcsname\relax
  \def\citenamefont#1{#1}\fi
\expandafter\ifx\csname url\endcsname\relax
  \def\url#1{\texttt{#1}}\fi
\expandafter\ifx\csname urlprefix\endcsname\relax\def\urlprefix{URL }\fi
\providecommand{\bibinfo}[2]{#2}
\providecommand{\eprint}[2][]{\url{#2}}

\bibitem[{\citenamefont{Matsukura et~al.}(2002)\citenamefont{Matsukura, Ohno,
  and Dietl}}]{Matsukura:2002_B}
\bibinfo{author}{\bibfnamefont{F.}~\bibnamefont{Matsukura}},
  \bibinfo{author}{\bibfnamefont{H.}~\bibnamefont{Ohno}}, \bibnamefont{and}
  \bibinfo{author}{\bibfnamefont{T.}~\bibnamefont{Dietl}}, in
  \emph{\bibinfo{booktitle}{III-V Ferromagnetic Semiconductors}}, edited by
  \bibinfo{editor}{\bibfnamefont{K.}~\bibnamefont{Buschow}}
  (\bibinfo{publisher}{Elsevier}, \bibinfo{year}{2002}),
  vol.~\bibinfo{volume}{14} of \emph{\bibinfo{series}{Handbook of Magnetic
  Materials}}.

\bibitem[{\citenamefont{Jungwirth et~al.}(2006)\citenamefont{Jungwirth, Sinova,
  Ma\v{s}ek, Ku\v{c}era, and MacDonald}}]{Jungwirth:2006_RMP}
\bibinfo{author}{\bibfnamefont{T.}~\bibnamefont{Jungwirth}},
  \bibinfo{author}{\bibfnamefont{J.}~\bibnamefont{Sinova}},
  \bibinfo{author}{\bibfnamefont{J.}~\bibnamefont{Ma\v{s}ek}},
  \bibinfo{author}{\bibfnamefont{J.}~\bibnamefont{Ku\v{c}era}},
  \bibnamefont{and} \bibinfo{author}{\bibfnamefont{A.~H.}
  \bibnamefont{MacDonald}}, \bibinfo{journal}{Rev. Mod. Phys.}
  \textbf{\bibinfo{volume}{78}}, \bibinfo{eid}{809} (\bibinfo{year}{2006}).

\bibitem[{\citenamefont{Dietl et~al.}(2007)\citenamefont{Dietl, Ohno, and
  Matsukura}}]{Dietl:2007_IEEE}
\bibinfo{author}{\bibfnamefont{T.}~\bibnamefont{Dietl}},
  \bibinfo{author}{\bibfnamefont{H.}~\bibnamefont{Ohno}}, \bibnamefont{and}
  \bibinfo{author}{\bibfnamefont{F.}~\bibnamefont{Matsukura}},
  \bibinfo{journal}{IEEE-Trans. Electronic Devices}
  \textbf{\bibinfo{volume}{54}}, \bibinfo{pages}{945} (\bibinfo{year}{2007}).

\bibitem[{\citenamefont{Farle}(1998)}]{Farle:1998_RPP}
\bibinfo{author}{\bibfnamefont{M.}~\bibnamefont{Farle}}, \bibinfo{journal}{Rep.
  Prog. Phys.} \textbf{\bibinfo{volume}{61}}, \bibinfo{pages}{755}
  (\bibinfo{year}{1998}).

\bibitem[{\citenamefont{Skomski}(2008)}]{Skomski:2008_B}
\bibinfo{author}{\bibfnamefont{R.}~\bibnamefont{Skomski}},
  \emph{\bibinfo{title}{Simple Models of Magnetism}}
  (\bibinfo{publisher}{Oxford New York}, \bibinfo{year}{2008}).

\bibitem[{\citenamefont{Kittel}(1987)}]{Kittel:1987_B}
\bibinfo{author}{\bibfnamefont{C.}~\bibnamefont{Kittel}},
  \emph{\bibinfo{title}{Quantum Theory of Solids}} (\bibinfo{publisher}{Wiley},
  \bibinfo{year}{1987}).

\bibitem[{\citenamefont{Timm and MacDonald}(2005)}]{Timm:2005_PRB}
\bibinfo{author}{\bibfnamefont{C.}~\bibnamefont{Timm}} \bibnamefont{and}
  \bibinfo{author}{\bibfnamefont{A.~H.} \bibnamefont{MacDonald}},
  \bibinfo{journal}{Phys. Rev. B} \textbf{\bibinfo{volume}{71}},
  \bibinfo{pages}{155206} (\bibinfo{year}{2005}).

\bibitem[{\citenamefont{Dzyaloshinskii}(1958)}]{Dzyaloshinskii:1958_JPCS}
\bibinfo{author}{\bibfnamefont{I.~E.} \bibnamefont{Dzyaloshinskii}},
  \bibinfo{journal}{J. Phys. Chem. Solids} \textbf{\bibinfo{volume}{4}},
  \bibinfo{pages}{241} (\bibinfo{year}{1958}).

\bibitem[{\citenamefont{Moriya}(1960)}]{Moriya:1960_PR}
\bibinfo{author}{\bibfnamefont{T.}~\bibnamefont{Moriya}},
  \bibinfo{journal}{Phys. Rev.} \textbf{\bibinfo{volume}{120}},
  \bibinfo{pages}{91} (\bibinfo{year}{1960}).

\bibitem[{\citenamefont{Konig et~al.}(2001)\citenamefont{Konig, Jungwirth, and
  MacDonald}}]{Konig:2001_PRB}
\bibinfo{author}{\bibfnamefont{J.}~\bibnamefont{Konig}},
  \bibinfo{author}{\bibfnamefont{T.}~\bibnamefont{Jungwirth}},
  \bibnamefont{and} \bibinfo{author}{\bibfnamefont{A.~H.}
  \bibnamefont{MacDonald}}, \bibinfo{journal}{Phys. Rev. B}
  \textbf{\bibinfo{volume}{64}}, \bibinfo{pages}{184423}
  (\bibinfo{year}{2001}).

\bibitem[{\citenamefont{Brey and G\'omez-Santos}(2003)}]{Brey:2003_PRB}
\bibinfo{author}{\bibfnamefont{L.}~\bibnamefont{Brey}} \bibnamefont{and}
  \bibinfo{author}{\bibfnamefont{G.}~\bibnamefont{G\'omez-Santos}},
  \bibinfo{journal}{Phys. Rev. B} \textbf{\bibinfo{volume}{68}},
  \bibinfo{pages}{115206} (\bibinfo{year}{2003}).

\bibitem[{\citenamefont{Dietl et~al.}(2000)\citenamefont{Dietl, Ohno,
  Matsukura, Cibert, and Ferrand}}]{Dietl:2000_S}
\bibinfo{author}{\bibfnamefont{T.}~\bibnamefont{Dietl}},
  \bibinfo{author}{\bibfnamefont{H.}~\bibnamefont{Ohno}},
  \bibinfo{author}{\bibfnamefont{F.}~\bibnamefont{Matsukura}},
  \bibinfo{author}{\bibfnamefont{J.}~\bibnamefont{Cibert}}, \bibnamefont{and}
  \bibinfo{author}{\bibfnamefont{D.}~\bibnamefont{Ferrand}},
  \bibinfo{journal}{Science} \textbf{\bibinfo{volume}{287}},
  \bibinfo{pages}{1019} (\bibinfo{year}{2000}).

\bibitem[{\citenamefont{Dietl et~al.}(2001{\natexlab{a}})\citenamefont{Dietl,
  Ohno, and Matsukura}}]{Dietl:2001_PRB}
\bibinfo{author}{\bibfnamefont{T.}~\bibnamefont{Dietl}},
  \bibinfo{author}{\bibfnamefont{H.}~\bibnamefont{Ohno}}, \bibnamefont{and}
  \bibinfo{author}{\bibfnamefont{F.}~\bibnamefont{Matsukura}},
  \bibinfo{journal}{Phys. Rev. B} \textbf{\bibinfo{volume}{63}},
  \bibinfo{pages}{195205} (\bibinfo{year}{2001}{\natexlab{a}}).

\bibitem[{\citenamefont{Abolfath et~al.}(2001)\citenamefont{Abolfath,
  Jungwirth, Brum, and MacDonald}}]{Abolfath:2001_PRB}
\bibinfo{author}{\bibfnamefont{M.}~\bibnamefont{Abolfath}},
  \bibinfo{author}{\bibfnamefont{T.}~\bibnamefont{Jungwirth}},
  \bibinfo{author}{\bibfnamefont{J.}~\bibnamefont{Brum}}, \bibnamefont{and}
  \bibinfo{author}{\bibfnamefont{A.~H.} \bibnamefont{MacDonald}},
  \bibinfo{journal}{Phys. Rev. B} \textbf{\bibinfo{volume}{63}},
  \bibinfo{pages}{054418} (\bibinfo{year}{2001}).

\bibitem[{\citenamefont{Dietl et~al.}(2001{\natexlab{b}})\citenamefont{Dietl,
  K\"onig, and MacDonald}}]{Dietl:2001_PRBb}
\bibinfo{author}{\bibfnamefont{T.}~\bibnamefont{Dietl}},
  \bibinfo{author}{\bibfnamefont{J.}~\bibnamefont{K\"onig}}, \bibnamefont{and}
  \bibinfo{author}{\bibfnamefont{A.~H.} \bibnamefont{MacDonald}},
  \bibinfo{journal}{Phys. Rev. B} \textbf{\bibinfo{volume}{64}},
  \bibinfo{pages}{241201} (\bibinfo{year}{2001}{\natexlab{b}}).

\bibitem[{\citenamefont{Bouzerar}(2007)}]{Bouzerar:2007_EPL}
\bibinfo{author}{\bibfnamefont{G.}~\bibnamefont{Bouzerar}},
  \bibinfo{journal}{Europhys. Lett.} \textbf{\bibinfo{volume}{79}},
  \bibinfo{pages}{57007} (\bibinfo{year}{2007}).

\bibitem[{\citenamefont{Potashnik et~al.}(2002)\citenamefont{Potashnik, Ku,
  Mahendiran, Chun, Wang, Samarth, and Schiffer}}]{Potashnik:2002_PRB}
\bibinfo{author}{\bibfnamefont{S.~J.} \bibnamefont{Potashnik}},
  \bibinfo{author}{\bibfnamefont{K.~C.} \bibnamefont{Ku}},
  \bibinfo{author}{\bibfnamefont{R.}~\bibnamefont{Mahendiran}},
  \bibinfo{author}{\bibfnamefont{S.~H.} \bibnamefont{Chun}},
  \bibinfo{author}{\bibfnamefont{R.~F.} \bibnamefont{Wang}},
  \bibinfo{author}{\bibfnamefont{N.}~\bibnamefont{Samarth}}, \bibnamefont{and}
  \bibinfo{author}{\bibfnamefont{P.}~\bibnamefont{Schiffer}},
  \bibinfo{journal}{Phys. Rev. B} \textbf{\bibinfo{volume}{66}},
  \bibinfo{pages}{012408} (\bibinfo{year}{2002}).

\bibitem[{\citenamefont{Wang et~al.}(2007)\citenamefont{Wang, Ren, Liu,
  Furdyna, Grimsditch, and Merlin}}]{Wang:2007_PRB}
\bibinfo{author}{\bibfnamefont{D.~M.} \bibnamefont{Wang}},
  \bibinfo{author}{\bibfnamefont{Y.~H.} \bibnamefont{Ren}},
  \bibinfo{author}{\bibfnamefont{X.}~\bibnamefont{Liu}},
  \bibinfo{author}{\bibfnamefont{J.~K.} \bibnamefont{Furdyna}},
  \bibinfo{author}{\bibfnamefont{M.}~\bibnamefont{Grimsditch}},
  \bibnamefont{and} \bibinfo{author}{\bibfnamefont{R.}~\bibnamefont{Merlin}},
  \bibinfo{journal}{Phys. Rev. B} \textbf{\bibinfo{volume}{75}},
  \bibinfo{eid}{233308} (\bibinfo{year}{2007}).

\bibitem[{\citenamefont{Zhou et~al.}(2007)\citenamefont{Zhou, Cho, Ge, Liu,
  Dobrowolska, and Furdyna}}]{Zhou:2007_IEEE}
\bibinfo{author}{\bibfnamefont{Y.}~\bibnamefont{Zhou}},
  \bibinfo{author}{\bibfnamefont{Y.}~\bibnamefont{Cho}},
  \bibinfo{author}{\bibfnamefont{Z.}~\bibnamefont{Ge}},
  \bibinfo{author}{\bibfnamefont{X.}~\bibnamefont{Liu}},
  \bibinfo{author}{\bibfnamefont{M.}~\bibnamefont{Dobrowolska}},
  \bibnamefont{and} \bibinfo{author}{\bibfnamefont{J.~K.}
  \bibnamefont{Furdyna}}, \bibinfo{journal}{IEEE Trans. Magn.}
  \textbf{\bibinfo{volume}{43}}, \bibinfo{pages}{3019} (\bibinfo{year}{2007}).

\bibitem[{\citenamefont{Liu et~al.}(2007)\citenamefont{Liu, Zhou, and
  Furdyna}}]{Liu:2007_PRB}
\bibinfo{author}{\bibfnamefont{X.}~\bibnamefont{Liu}},
  \bibinfo{author}{\bibfnamefont{Y.~Y.} \bibnamefont{Zhou}}, \bibnamefont{and}
  \bibinfo{author}{\bibfnamefont{J.~K.} \bibnamefont{Furdyna}},
  \bibinfo{journal}{Phys. Rev. B} \textbf{\bibinfo{volume}{75}},
  \bibinfo{pages}{195220} (\bibinfo{year}{2007}).

\bibitem[{\citenamefont{Bihler et~al.}(2009)\citenamefont{Bihler, Schoch,
  Limmer, Goennenwein, and Brandt}}]{Bihler:2009_PRB}
\bibinfo{author}{\bibfnamefont{C.}~\bibnamefont{Bihler}},
  \bibinfo{author}{\bibfnamefont{W.}~\bibnamefont{Schoch}},
  \bibinfo{author}{\bibfnamefont{W.}~\bibnamefont{Limmer}},
  \bibinfo{author}{\bibfnamefont{S.~T.~B.} \bibnamefont{Goennenwein}},
  \bibnamefont{and} \bibinfo{author}{\bibfnamefont{M.~S.}
  \bibnamefont{Brandt}}, \bibinfo{journal}{Phys. Rev. B}
  \textbf{\bibinfo{volume}{79}}, \bibinfo{pages}{045205}
  (\bibinfo{year}{2009}).

\bibitem[{\citenamefont{Gourdon et~al.}(2007)\citenamefont{Gourdon, Dourlat,
  Jeudy, Khazen, von Bardeleben, Thevenard, and
  Lema\^{i}tre}}]{Gourdon:2007_PRB}
\bibinfo{author}{\bibfnamefont{C.}~\bibnamefont{Gourdon}},
  \bibinfo{author}{\bibfnamefont{A.}~\bibnamefont{Dourlat}},
  \bibinfo{author}{\bibfnamefont{V.}~\bibnamefont{Jeudy}},
  \bibinfo{author}{\bibfnamefont{K.}~\bibnamefont{Khazen}},
  \bibinfo{author}{\bibfnamefont{H.~J.} \bibnamefont{von Bardeleben}},
  \bibinfo{author}{\bibfnamefont{L.}~\bibnamefont{Thevenard}},
  \bibnamefont{and}
  \bibinfo{author}{\bibfnamefont{A.}~\bibnamefont{Lema\^{i}tre}},
  \bibinfo{journal}{Phys. Rev. B} \textbf{\bibinfo{volume}{76}},
  \bibinfo{eid}{241301} (\bibinfo{year}{2007}).

\bibitem[{\citenamefont{Rader et~al.}(2004)\citenamefont{Rader, Pampuch,
  Shikin, Gudat, Okabayashi, Mizokawa, Fujimori, Hayashi, Tanaka, Tanaka
  et~al.}}]{Rader:2004_PRB}
\bibinfo{author}{\bibfnamefont{O.}~\bibnamefont{Rader}},
  \bibinfo{author}{\bibfnamefont{C.}~\bibnamefont{Pampuch}},
  \bibinfo{author}{\bibfnamefont{A.~M.} \bibnamefont{Shikin}},
  \bibinfo{author}{\bibfnamefont{W.}~\bibnamefont{Gudat}},
  \bibinfo{author}{\bibfnamefont{J.}~\bibnamefont{Okabayashi}},
  \bibinfo{author}{\bibfnamefont{T.}~\bibnamefont{Mizokawa}},
  \bibinfo{author}{\bibfnamefont{A.}~\bibnamefont{Fujimori}},
  \bibinfo{author}{\bibfnamefont{T.}~\bibnamefont{Hayashi}},
  \bibinfo{author}{\bibfnamefont{M.}~\bibnamefont{Tanaka}},
  \bibinfo{author}{\bibfnamefont{A.}~\bibnamefont{Tanaka}},
  \bibnamefont{et~al.}, \bibinfo{journal}{Phys. Rev. B}
  \textbf{\bibinfo{volume}{69}}, \bibinfo{pages}{075202}
  (\bibinfo{year}{2004}).

\bibitem[{\citenamefont{Schulthess et~al.}(2005)\citenamefont{Schulthess,
  Temmerman, Szotek, Butler, and Stocks}}]{Schulthess:2005_NM}
\bibinfo{author}{\bibfnamefont{T.~C.} \bibnamefont{Schulthess}},
  \bibinfo{author}{\bibfnamefont{W.~M.} \bibnamefont{Temmerman}},
  \bibinfo{author}{\bibfnamefont{Z.}~\bibnamefont{Szotek}},
  \bibinfo{author}{\bibfnamefont{W.~H.} \bibnamefont{Butler}},
  \bibnamefont{and} \bibinfo{author}{\bibfnamefont{G.~M.}
  \bibnamefont{Stocks}}, \bibinfo{journal}{Nature Mat.}
  \textbf{\bibinfo{volume}{4}}, \bibinfo{pages}{838} (\bibinfo{year}{2005}).

\bibitem[{\citenamefont{Dietl}(2008{\natexlab{a}})}]{Dietl:2008_JPSJ}
\bibinfo{author}{\bibfnamefont{T.}~\bibnamefont{Dietl}}, \bibinfo{journal}{J.
  Phys. Soc. Jap.} \textbf{\bibinfo{volume}{77}}, \bibinfo{pages}{031005}
  (\bibinfo{year}{2008}{\natexlab{a}}).

\bibitem[{\citenamefont{Neumaier et~al.}(2009)\citenamefont{Neumaier, Turek,
  Wurstbauer, Vogl, Utz, Wegscheider, and Weiss}}]{Neumaier:2009_PRL}
\bibinfo{author}{\bibfnamefont{D.}~\bibnamefont{Neumaier}},
  \bibinfo{author}{\bibfnamefont{M.}~\bibnamefont{Turek}},
  \bibinfo{author}{\bibfnamefont{U.}~\bibnamefont{Wurstbauer}},
  \bibinfo{author}{\bibfnamefont{A.}~\bibnamefont{Vogl}},
  \bibinfo{author}{\bibfnamefont{M.}~\bibnamefont{Utz}},
  \bibinfo{author}{\bibfnamefont{W.}~\bibnamefont{Wegscheider}},
  \bibnamefont{and} \bibinfo{author}{\bibfnamefont{D.}~\bibnamefont{Weiss}},
  \bibinfo{journal}{Phys. Rev. Lett.} \textbf{\bibinfo{volume}{103}},
  \bibinfo{pages}{087203} (\bibinfo{year}{2009}).

\bibitem[{\citenamefont{Jancu et~al.}(1998)\citenamefont{Jancu, Scholz,
  Beltram, and Bassani}}]{Jancu:1998_PRB}
\bibinfo{author}{\bibfnamefont{J.-M.} \bibnamefont{Jancu}},
  \bibinfo{author}{\bibfnamefont{R.}~\bibnamefont{Scholz}},
  \bibinfo{author}{\bibfnamefont{F.}~\bibnamefont{Beltram}}, \bibnamefont{and}
  \bibinfo{author}{\bibfnamefont{F.}~\bibnamefont{Bassani}},
  \bibinfo{journal}{Phys. Rev. B} \textbf{\bibinfo{volume}{57}},
  \bibinfo{pages}{6493} (\bibinfo{year}{1998}).

\bibitem[{\citenamefont{Strahberger and Vogl}(2000)}]{Strahberger:2000_PRB}
\bibinfo{author}{\bibfnamefont{C.}~\bibnamefont{Strahberger}} \bibnamefont{and}
  \bibinfo{author}{\bibfnamefont{P.}~\bibnamefont{Vogl}},
  \bibinfo{journal}{Phys. Rev. B} \textbf{\bibinfo{volume}{62}},
  \bibinfo{pages}{7289} (\bibinfo{year}{2000}).

\bibitem[{\citenamefont{Sankowski et~al.}(2007)\citenamefont{Sankowski, Kacman,
  Majewski, and Dietl}}]{Sankowski:2007_PRB}
\bibinfo{author}{\bibfnamefont{P.}~\bibnamefont{Sankowski}},
  \bibinfo{author}{\bibfnamefont{P.}~\bibnamefont{Kacman}},
  \bibinfo{author}{\bibfnamefont{J.~A.} \bibnamefont{Majewski}},
  \bibnamefont{and} \bibinfo{author}{\bibfnamefont{T.}~\bibnamefont{Dietl}},
  \bibinfo{journal}{Phys. Rev. B} \textbf{\bibinfo{volume}{75}},
  \bibinfo{eid}{045306} (\bibinfo{year}{2007}).

\bibitem[{\citenamefont{Oszwa\l{}dowski
  et~al.}(2006)\citenamefont{Oszwa\l{}dowski, Majewski, and
  Dietl}}]{Oszwaldowski:2006_PRB}
\bibinfo{author}{\bibfnamefont{R.}~\bibnamefont{Oszwa\l{}dowski}},
  \bibinfo{author}{\bibfnamefont{J.~A.} \bibnamefont{Majewski}},
  \bibnamefont{and} \bibinfo{author}{\bibfnamefont{T.}~\bibnamefont{Dietl}},
  \bibinfo{journal}{Phys. Rev. B} \textbf{\bibinfo{volume}{74}},
  \bibinfo{eid}{153310} (\bibinfo{year}{2006}).

\bibitem[{\citenamefont{Werpachowska and Dietl}(2010)}]{Werpachowska:2010_PRB}
\bibinfo{author}{\bibfnamefont{A.}~\bibnamefont{Werpachowska}}
  \bibnamefont{and} \bibinfo{author}{\bibfnamefont{T.}~\bibnamefont{Dietl}},
  \bibinfo{journal}{Phys. Rev. B} \textbf{\bibinfo{volume}{81}},
  \bibinfo{pages}{155205} (\bibinfo{year}{2010}).

\bibitem[{\citenamefont{Chiba et~al.}(2010)\citenamefont{Chiba, Werpachowska,
  Endo, Nishitani, Matsukura, Dietl, and Ohno}}]{Chiba:2010_PRL}
\bibinfo{author}{\bibfnamefont{D.}~\bibnamefont{Chiba}},
  \bibinfo{author}{\bibfnamefont{A.}~\bibnamefont{Werpachowska}},
  \bibinfo{author}{\bibfnamefont{M.}~\bibnamefont{Endo}},
  \bibinfo{author}{\bibfnamefont{Y.}~\bibnamefont{Nishitani}},
  \bibinfo{author}{\bibfnamefont{F.}~\bibnamefont{Matsukura}},
  \bibinfo{author}{\bibfnamefont{T.}~\bibnamefont{Dietl}}, \bibnamefont{and}
  \bibinfo{author}{\bibfnamefont{H.}~\bibnamefont{Ohno}},
  \bibinfo{journal}{Phys. Rev. Lett.} \textbf{\bibinfo{volume}{104}},
  \bibinfo{pages}{106601} (\bibinfo{year}{2010}).

\bibitem[{\citenamefont{Dresselhaus}(1955)}]{Dresselhaus:1955_PR}
\bibinfo{author}{\bibfnamefont{G.}~\bibnamefont{Dresselhaus}},
  \bibinfo{journal}{Phys. Rev.} \textbf{\bibinfo{volume}{100}},
  \bibinfo{pages}{580} (\bibinfo{year}{1955}).

\bibitem[{\citenamefont{Bychkov and Rashba}(1984)}]{Bychkov:1984_JPh}
\bibinfo{author}{\bibfnamefont{Y.~L.} \bibnamefont{Bychkov}} \bibnamefont{and}
  \bibinfo{author}{\bibfnamefont{E.~I.} \bibnamefont{Rashba}},
  \bibinfo{journal}{J. Phys.} \textbf{\bibinfo{volume}{C 17}},
  \bibinfo{pages}{6093} (\bibinfo{year}{1984}).

\bibitem[{\citenamefont{Okabayashi et~al.}(2002)\citenamefont{Okabayashi,
  Mizokawa, Sarma, Fujimori, Slupinski, Oiwa, and
  Munekata}}]{PhysRevB.65.161203}
\bibinfo{author}{\bibfnamefont{J.}~\bibnamefont{Okabayashi}},
  \bibinfo{author}{\bibfnamefont{T.}~\bibnamefont{Mizokawa}},
  \bibinfo{author}{\bibfnamefont{D.~D.} \bibnamefont{Sarma}},
  \bibinfo{author}{\bibfnamefont{A.}~\bibnamefont{Fujimori}},
  \bibinfo{author}{\bibfnamefont{T.}~\bibnamefont{Slupinski}},
  \bibinfo{author}{\bibfnamefont{A.}~\bibnamefont{Oiwa}}, \bibnamefont{and}
  \bibinfo{author}{\bibfnamefont{H.}~\bibnamefont{Munekata}},
  \bibinfo{journal}{Phys. Rev. B} \textbf{\bibinfo{volume}{65}},
  \bibinfo{pages}{161203} (\bibinfo{year}{2002}).

\bibitem[{\citenamefont{Dietl}(2008{\natexlab{b}})}]{Dietl:2008_PRB}
\bibinfo{author}{\bibfnamefont{T.}~\bibnamefont{Dietl}},
  \bibinfo{journal}{Phys. Rev. B} \textbf{\bibinfo{volume}{77}},
  \bibinfo{pages}{085208} (\bibinfo{year}{2008}{\natexlab{b}}).

\bibitem[{\citenamefont{L\"owdin}(1951)}]{Loewdin:1951_JCP}
\bibinfo{author}{\bibfnamefont{P.}~\bibnamefont{L\"owdin}},
  \bibinfo{journal}{J. Chem. Phys.} \textbf{\bibinfo{volume}{19}},
  \bibinfo{pages}{1396} (\bibinfo{year}{1951}).

\bibitem[{\citenamefont{Thijssen}(2007)}]{Thijssen:2007_B}
\bibinfo{author}{\bibfnamefont{J.~M.} \bibnamefont{Thijssen}},
  \emph{\bibinfo{title}{Computational Physics}} (\bibinfo{publisher}{Cambridge
  University Press}, \bibinfo{year}{2007}).

\bibitem[{\citenamefont{Ziener et~al.}(2004)\citenamefont{Ziener, Glutsch, and
  Bechstedt}}]{Ziener:2004_PRB}
\bibinfo{author}{\bibfnamefont{C.~H.} \bibnamefont{Ziener}},
  \bibinfo{author}{\bibfnamefont{S.}~\bibnamefont{Glutsch}}, \bibnamefont{and}
  \bibinfo{author}{\bibfnamefont{F.}~\bibnamefont{Bechstedt}},
  \bibinfo{journal}{Phys. Rev. B} \textbf{\bibinfo{volume}{70}},
  \bibinfo{pages}{075205} (\bibinfo{year}{2004}).

\bibitem[{\citenamefont{Werpachowska and
  Wilamowski}(2006)}]{Werpachowska:2006_MS}
\bibinfo{author}{\bibfnamefont{A.}~\bibnamefont{Werpachowska}}
  \bibnamefont{and}
  \bibinfo{author}{\bibfnamefont{Z.}~\bibnamefont{Wilamowski}},
  \bibinfo{journal}{Mat. Sci.-PL} \textbf{\bibinfo{volume}{24}},
  \bibinfo{pages}{675} (\bibinfo{year}{2006}).

\bibitem[{\citenamefont{Holstein and Primakoff}(1940)}]{Holstein:1940_PR}
\bibinfo{author}{\bibfnamefont{T.}~\bibnamefont{Holstein}} \bibnamefont{and}
  \bibinfo{author}{\bibfnamefont{H.}~\bibnamefont{Primakoff}},
  \bibinfo{journal}{Phys. Rev.} \textbf{\bibinfo{volume}{58}},
  \bibinfo{pages}{1098} (\bibinfo{year}{1940}).

\bibitem[{\citenamefont{Chen et~al.}(1985)\citenamefont{Chen, Dobrowolska,
  Furdyna, and Rodriguez}}]{Chen:1985_PRB}
\bibinfo{author}{\bibfnamefont{Y.-F.} \bibnamefont{Chen}},
  \bibinfo{author}{\bibfnamefont{M.}~\bibnamefont{Dobrowolska}},
  \bibinfo{author}{\bibfnamefont{J.~K.} \bibnamefont{Furdyna}},
  \bibnamefont{and}
  \bibinfo{author}{\bibfnamefont{S.}~\bibnamefont{Rodriguez}},
  \bibinfo{journal}{Phys. Rev. B} \textbf{\bibinfo{volume}{32}},
  \bibinfo{pages}{890} (\bibinfo{year}{1985}).

\bibitem[{\citenamefont{Pappert et~al.}(2007)\citenamefont{Pappert, Humpfner,
  Wenisch, Brunner, Gould, Schmidt, and Molenkamp}}]{Pappert:2007_APL}
\bibinfo{author}{\bibfnamefont{K.}~\bibnamefont{Pappert}},
  \bibinfo{author}{\bibfnamefont{S.}~\bibnamefont{Humpfner}},
  \bibinfo{author}{\bibfnamefont{J.}~\bibnamefont{Wenisch}},
  \bibinfo{author}{\bibfnamefont{K.}~\bibnamefont{Brunner}},
  \bibinfo{author}{\bibfnamefont{C.}~\bibnamefont{Gould}},
  \bibinfo{author}{\bibfnamefont{G.}~\bibnamefont{Schmidt}}, \bibnamefont{and}
  \bibinfo{author}{\bibfnamefont{L.~W.} \bibnamefont{Molenkamp}},
  \bibinfo{journal}{Appl. Phys. Lett.} \textbf{\bibinfo{volume}{90}},
  \bibinfo{eid}{062109} (\bibinfo{year}{2007}).

\bibitem[{\citenamefont{Dietl et~al.}(1999)\citenamefont{Dietl, Cibert,
  Ferrand, and {Merle d'Aubign\'e}}}]{Dietl:1999_MSE}
\bibinfo{author}{\bibfnamefont{T.}~\bibnamefont{Dietl}},
  \bibinfo{author}{\bibfnamefont{J.}~\bibnamefont{Cibert}},
  \bibinfo{author}{\bibfnamefont{D.}~\bibnamefont{Ferrand}}, \bibnamefont{and}
  \bibinfo{author}{\bibfnamefont{Y.}~\bibnamefont{{Merle d'Aubign\'e}}},
  \bibinfo{journal}{Materials Sci. Engin.} \textbf{\bibinfo{volume}{63}},
  \bibinfo{pages}{103} (\bibinfo{year}{1999}).

\bibitem[{\citenamefont{Ruderman and Kittel}(1954)}]{Ruderman:1954_PR}
\bibinfo{author}{\bibfnamefont{M.~A.} \bibnamefont{Ruderman}} \bibnamefont{and}
  \bibinfo{author}{\bibfnamefont{C.}~\bibnamefont{Kittel}},
  \bibinfo{journal}{Phys. Rev.} \textbf{\bibinfo{volume}{96}},
  \bibinfo{pages}{99} (\bibinfo{year}{1954}).

\bibitem[{\citenamefont{Broerman et~al.}(1971)\citenamefont{Broerman, Liu, and
  Pathak}}]{Broerman:1971_PRB}
\bibinfo{author}{\bibfnamefont{J.~G.} \bibnamefont{Broerman}},
  \bibinfo{author}{\bibfnamefont{L.}~\bibnamefont{Liu}}, \bibnamefont{and}
  \bibinfo{author}{\bibfnamefont{K.~N.} \bibnamefont{Pathak}},
  \bibinfo{journal}{Phys. Rev. B} \textbf{\bibinfo{volume}{4}},
  \bibinfo{pages}{664} (\bibinfo{year}{1971}).

\bibitem[{\citenamefont{{Szyma\'nska} and Dietl}(1978)}]{Szymanska:1978_JPCS}
\bibinfo{author}{\bibfnamefont{W.}~\bibnamefont{{Szyma\'nska}}}
  \bibnamefont{and} \bibinfo{author}{\bibfnamefont{T.}~\bibnamefont{Dietl}},
  \bibinfo{journal}{J. Phys. Chem. Solids} \textbf{\bibinfo{volume}{39}},
  \bibinfo{pages}{1025} (\bibinfo{year}{1978}).

\bibitem[{\citenamefont{Binney et~al.}(1992)\citenamefont{Binney, Fisher, and
  J.}}]{Binney:1992_B}
\bibinfo{author}{\bibfnamefont{J.~J.} \bibnamefont{Binney}},
  \bibinfo{author}{\bibfnamefont{A.~J.} \bibnamefont{Fisher}},
  \bibnamefont{and} \bibinfo{author}{\bibfnamefont{N.~M.~E.} \bibnamefont{J.}},
  \emph{\bibinfo{title}{The Theory of Critical Phenomena}}
  (\bibinfo{publisher}{Oxford University Press}, \bibinfo{year}{1992}).

\bibitem[{\citenamefont{K\"onig et~al.}(2000)\citenamefont{K\"onig, Lin, and
  MacDonald}}]{Konig:2000_PRL}
\bibinfo{author}{\bibfnamefont{J.}~\bibnamefont{K\"onig}},
  \bibinfo{author}{\bibfnamefont{H.-H.} \bibnamefont{Lin}}, \bibnamefont{and}
  \bibinfo{author}{\bibfnamefont{A.~H.} \bibnamefont{MacDonald}},
  \bibinfo{journal}{Phys. Rev. Lett.} \textbf{\bibinfo{volume}{84}},
  \bibinfo{pages}{5628} (\bibinfo{year}{2000}).

\bibitem[{\citenamefont{Callen}(1963)}]{Callen:1963_PR}
\bibinfo{author}{\bibfnamefont{H.~B.} \bibnamefont{Callen}},
  \bibinfo{journal}{Phys. Rev.} \textbf{\bibinfo{volume}{130}},
  \bibinfo{pages}{890} (\bibinfo{year}{1963}).

\bibitem[{\citenamefont{Yang et~al.}(2001)\citenamefont{Yang, Sun, and
  Chang}}]{Yang:2001_PRL}
\bibinfo{author}{\bibfnamefont{M.-F.} \bibnamefont{Yang}},
  \bibinfo{author}{\bibfnamefont{S.-J.} \bibnamefont{Sun}}, \bibnamefont{and}
  \bibinfo{author}{\bibfnamefont{M.-C.} \bibnamefont{Chang}},
  \bibinfo{journal}{Phys. Rev. Lett.} \textbf{\bibinfo{volume}{86}},
  \bibinfo{pages}{5636} (\bibinfo{year}{2001}).

\bibitem[{\citenamefont{Berciu and Bhatt}(2002)}]{Berciu:2002_PRB}
\bibinfo{author}{\bibfnamefont{M.}~\bibnamefont{Berciu}} \bibnamefont{and}
  \bibinfo{author}{\bibfnamefont{R.~N.} \bibnamefont{Bhatt}},
  \bibinfo{journal}{Phys. Rev. B} \textbf{\bibinfo{volume}{66}},
  \bibinfo{pages}{085207} (\bibinfo{year}{2002}).

\bibitem[{\citenamefont{Meilikhov}(2007)}]{Meilikhov:2007_PRB}
\bibinfo{author}{\bibfnamefont{E.~Z.} \bibnamefont{Meilikhov}},
  \bibinfo{journal}{Phys. Rev. B} \textbf{\bibinfo{volume}{75}},
  \bibinfo{pages}{045204} (\bibinfo{year}{2007}).

\bibitem[{\citenamefont{Singh et~al.}(2007)\citenamefont{Singh, Das, Sharma,
  and Nolting}}]{Singh:2007_JP}
\bibinfo{author}{\bibfnamefont{A.}~\bibnamefont{Singh}},
  \bibinfo{author}{\bibfnamefont{S.}~\bibnamefont{Das}},
  \bibinfo{author}{\bibfnamefont{A.}~\bibnamefont{Sharma}}, \bibnamefont{and}
  \bibinfo{author}{\bibfnamefont{A.}~\bibnamefont{Nolting}},
  \bibinfo{journal}{J. Phys.: Condens. Matter} \textbf{\bibinfo{volume}{19}},
  \bibinfo{pages}{236213} (\bibinfo{year}{2007}).

\bibitem[{\citenamefont{Sza\l{}owski and Balcerzak}(2008)}]{Szalowski:2008_PRB}
\bibinfo{author}{\bibfnamefont{K.}~\bibnamefont{Sza\l{}owski}}
  \bibnamefont{and}
  \bibinfo{author}{\bibfnamefont{T.}~\bibnamefont{Balcerzak}},
  \bibinfo{journal}{Phys. Rev. B} \textbf{\bibinfo{volume}{77}},
  \bibinfo{pages}{115204} (\bibinfo{year}{2008}).

\bibitem[{\citenamefont{Tang and Nolting}(2007)}]{Tang:2008_PRB}
\bibinfo{author}{\bibfnamefont{G.}~\bibnamefont{Tang}} \bibnamefont{and}
  \bibinfo{author}{\bibfnamefont{W.}~\bibnamefont{Nolting}},
  \bibinfo{journal}{Phys. Status Solidi B} \textbf{\bibinfo{volume}{244}},
  \bibinfo{pages}{735} (\bibinfo{year}{2007}).

\bibitem[{\citenamefont{Balcerzak and Sza\l{}owski}(2009)}]{Balcerzak:2009_PRB}
\bibinfo{author}{\bibfnamefont{T.}~\bibnamefont{Balcerzak}} \bibnamefont{and}
  \bibinfo{author}{\bibfnamefont{K.}~\bibnamefont{Sza\l{}owski}},
  \bibinfo{journal}{Phys. Rev. B} \textbf{\bibinfo{volume}{80}},
  \bibinfo{pages}{144404} (\bibinfo{year}{2009}).

\bibitem[{\citenamefont{Bloch}(1930)}]{Bloch:1930_ZP}
\bibinfo{author}{\bibfnamefont{F.}~\bibnamefont{Bloch}}, \bibinfo{journal}{Z.
  Phys.} \textbf{\bibinfo{volume}{61}}, \bibinfo{pages}{206}
  (\bibinfo{year}{1930}).

\bibitem[{\citenamefont{Mermin and Wagner}(1966)}]{Mermin:1966_PRL}
\bibinfo{author}{\bibfnamefont{N.~D.} \bibnamefont{Mermin}} \bibnamefont{and}
  \bibinfo{author}{\bibfnamefont{H.}~\bibnamefont{Wagner}},
  \bibinfo{journal}{Phys. Rev. Lett.} \textbf{\bibinfo{volume}{17}},
  \bibinfo{pages}{1133} (\bibinfo{year}{1966}).

\bibitem[{\citenamefont{Jackson et~al.}(1989)\citenamefont{Jackson, Liao,
  Bhagat, and Manheimer}}]{Jackson:1989_JMMM}
\bibinfo{author}{\bibfnamefont{E.}~\bibnamefont{Jackson}},
  \bibinfo{author}{\bibfnamefont{S.}~\bibnamefont{Liao}},
  \bibinfo{author}{\bibfnamefont{S.}~\bibnamefont{Bhagat}}, \bibnamefont{and}
  \bibinfo{author}{\bibfnamefont{M.}~\bibnamefont{Manheimer}},
  \bibinfo{journal}{J. Magn. Magn. Mater.} \textbf{\bibinfo{volume}{80}},
  \bibinfo{pages}{229 } (\bibinfo{year}{1989}).

\bibitem[{\citenamefont{Huang}(1987)}]{Huang:1987_B}
\bibinfo{author}{\bibfnamefont{K.}~\bibnamefont{Huang}},
  \emph{\bibinfo{title}{Statistical mechanics}} (\bibinfo{publisher}{Wiley, New
  York}, \bibinfo{year}{1987}).

\bibitem[{\citenamefont{Aharoni}(2001)}]{Aharoni:2001_B}
\bibinfo{author}{\bibfnamefont{A.}~\bibnamefont{Aharoni}},
  \emph{\bibinfo{title}{Introduction to the Theory of Ferromagnetism}}
  (\bibinfo{publisher}{Oxford University Press}, \bibinfo{year}{2001}).

\bibitem[{\citenamefont{\'{S}liwa and Dietl}(2006)}]{Sliwa:2006_PRB}
\bibinfo{author}{\bibfnamefont{C.}~\bibnamefont{\'{S}liwa}} \bibnamefont{and}
  \bibinfo{author}{\bibfnamefont{T.}~\bibnamefont{Dietl}},
  \bibinfo{journal}{Phys. Rev. B} \textbf{\bibinfo{volume}{74}},
  \bibinfo{eid}{245215} (\bibinfo{year}{2006}).

\bibitem[{\citenamefont{Blinowski and Kacman}(2003)}]{Blinowski:2002_PRB}
\bibinfo{author}{\bibfnamefont{J.}~\bibnamefont{Blinowski}} \bibnamefont{and}
  \bibinfo{author}{\bibfnamefont{P.}~\bibnamefont{Kacman}},
  \bibinfo{journal}{Phys. Rev. B} \textbf{\bibinfo{volume}{67}},
  \bibinfo{pages}{121204} (\bibinfo{year}{2003}).

\bibitem[{\citenamefont{Edmonds et~al.}(2004)\citenamefont{Edmonds,
  Bogus\l{}awski, Wang, Campion, Novikov, Farley, Gallagher, Foxon, Sawicki,
  Dietl et~al.}}]{Edmonds:2004_PRL}
\bibinfo{author}{\bibfnamefont{K.~W.} \bibnamefont{Edmonds}},
  \bibinfo{author}{\bibfnamefont{P.}~\bibnamefont{Bogus\l{}awski}},
  \bibinfo{author}{\bibfnamefont{K.~Y.} \bibnamefont{Wang}},
  \bibinfo{author}{\bibfnamefont{R.~P.} \bibnamefont{Campion}},
  \bibinfo{author}{\bibfnamefont{S.~N.} \bibnamefont{Novikov}},
  \bibinfo{author}{\bibfnamefont{N.~R.~S.} \bibnamefont{Farley}},
  \bibinfo{author}{\bibfnamefont{B.~L.} \bibnamefont{Gallagher}},
  \bibinfo{author}{\bibfnamefont{C.~T.} \bibnamefont{Foxon}},
  \bibinfo{author}{\bibfnamefont{M.}~\bibnamefont{Sawicki}},
  \bibinfo{author}{\bibfnamefont{T.}~\bibnamefont{Dietl}},
  \bibnamefont{et~al.}, \bibinfo{journal}{Phys. Rev. Lett.}
  \textbf{\bibinfo{volume}{92}}, \bibinfo{pages}{037201}
  (\bibinfo{year}{2004}).

\bibitem[{\citenamefont{Stefanowicz et~al.}(2010)\citenamefont{Stefanowicz,
  \'Sliwa, Aleshkevych, Dietl, D\"oppe, Wurstbauer, Wegscheider, Weiss, and
  Sawicki}}]{Stefanowicz:2010_PRB}
\bibinfo{author}{\bibfnamefont{W.}~\bibnamefont{Stefanowicz}},
  \bibinfo{author}{\bibfnamefont{C.}~\bibnamefont{\'Sliwa}},
  \bibinfo{author}{\bibfnamefont{P.}~\bibnamefont{Aleshkevych}},
  \bibinfo{author}{\bibfnamefont{T.}~\bibnamefont{Dietl}},
  \bibinfo{author}{\bibfnamefont{M.}~\bibnamefont{D\"oppe}},
  \bibinfo{author}{\bibfnamefont{U.}~\bibnamefont{Wurstbauer}},
  \bibinfo{author}{\bibfnamefont{W.}~\bibnamefont{Wegscheider}},
  \bibinfo{author}{\bibfnamefont{D.}~\bibnamefont{Weiss}}, \bibnamefont{and}
  \bibinfo{author}{\bibfnamefont{M.}~\bibnamefont{Sawicki}},
  \bibinfo{journal}{Phys. Rev. B} \textbf{\bibinfo{volume}{81}},
  \bibinfo{pages}{155203} (\bibinfo{year}{2010}).

\bibitem[{\citenamefont{Rappoport et~al.}(2004)\citenamefont{Rappoport,
  Redli\ifmmode~\acute{n}\else \'{n}\fi{}ski, Liu, Zar\'and, Furdyna, and
  Jank\'o}}]{Rappoport:2004_PRB}
\bibinfo{author}{\bibfnamefont{T.~G.} \bibnamefont{Rappoport}},
  \bibinfo{author}{\bibfnamefont{P.}~\bibnamefont{Redli\ifmmode~\acute{n}\else
  \'{n}\fi{}ski}}, \bibinfo{author}{\bibfnamefont{X.}~\bibnamefont{Liu}},
  \bibinfo{author}{\bibfnamefont{G.}~\bibnamefont{Zar\'and}},
  \bibinfo{author}{\bibfnamefont{J.~K.} \bibnamefont{Furdyna}},
  \bibnamefont{and} \bibinfo{author}{\bibfnamefont{B.}~\bibnamefont{Jank\'o}},
  \bibinfo{journal}{Phys. Rev. B} \textbf{\bibinfo{volume}{69}},
  \bibinfo{pages}{125213} (\bibinfo{year}{2004}).

\bibitem[{\citenamefont{Zar\'and and Jank\'o}(2002)}]{Zarand:2002_PRL}
\bibinfo{author}{\bibfnamefont{G.}~\bibnamefont{Zar\'and}} \bibnamefont{and}
  \bibinfo{author}{\bibfnamefont{B.}~\bibnamefont{Jank\'o}},
  \bibinfo{journal}{Phys. Rev. Lett.} \textbf{\bibinfo{volume}{89}},
  \bibinfo{pages}{047201} (\bibinfo{year}{2002}).

\bibitem[{\citenamefont{Sawicki et~al.}(2004)\citenamefont{Sawicki, Matsukura,
  Idziaszek, Dietl, Schott, Ruester, Gould, Karczewski, Schmidt, and
  Molenkamp}}]{Sawicki:2004_PRB}
\bibinfo{author}{\bibfnamefont{M.}~\bibnamefont{Sawicki}},
  \bibinfo{author}{\bibfnamefont{F.}~\bibnamefont{Matsukura}},
  \bibinfo{author}{\bibfnamefont{A.}~\bibnamefont{Idziaszek}},
  \bibinfo{author}{\bibfnamefont{T.}~\bibnamefont{Dietl}},
  \bibinfo{author}{\bibfnamefont{G.~M.} \bibnamefont{Schott}},
  \bibinfo{author}{\bibfnamefont{C.}~\bibnamefont{Ruester}},
  \bibinfo{author}{\bibfnamefont{C.}~\bibnamefont{Gould}},
  \bibinfo{author}{\bibfnamefont{G.}~\bibnamefont{Karczewski}},
  \bibinfo{author}{\bibfnamefont{G.}~\bibnamefont{Schmidt}}, \bibnamefont{and}
  \bibinfo{author}{\bibfnamefont{L.~W.} \bibnamefont{Molenkamp}},
  \bibinfo{journal}{Phys. Rev. B} \textbf{\bibinfo{volume}{70}},
  \bibinfo{pages}{245325} (\bibinfo{year}{2004}).

\bibitem[{\citenamefont{Welp et~al.}(2004)\citenamefont{Welp, Vlasko-Vlasov,
  Menzel, You, Liu, Furdyna, and Wojtowicz}}]{Welp:2004_APL}
\bibinfo{author}{\bibfnamefont{U.}~\bibnamefont{Welp}},
  \bibinfo{author}{\bibfnamefont{V.~K.} \bibnamefont{Vlasko-Vlasov}},
  \bibinfo{author}{\bibfnamefont{A.}~\bibnamefont{Menzel}},
  \bibinfo{author}{\bibfnamefont{H.~D.} \bibnamefont{You}},
  \bibinfo{author}{\bibfnamefont{X.}~\bibnamefont{Liu}},
  \bibinfo{author}{\bibfnamefont{J.~K.} \bibnamefont{Furdyna}},
  \bibnamefont{and}
  \bibinfo{author}{\bibfnamefont{T.}~\bibnamefont{Wojtowicz}},
  \bibinfo{journal}{Appl. Phys. Lett.} \textbf{\bibinfo{volume}{85}},
  \bibinfo{pages}{260} (\bibinfo{year}{2004}).

\bibitem[{\citenamefont{Sawicki et~al.}(2005)\citenamefont{Sawicki, Wang,
  Edmonds, Campion, Staddon, Farley, Foxon, Papis, Kami\ifmmode~\acute{n}\else
  \'{n}\fi{}ska, Piotrowska et~al.}}]{Sawicki:2005_PRB}
\bibinfo{author}{\bibfnamefont{M.}~\bibnamefont{Sawicki}},
  \bibinfo{author}{\bibfnamefont{K.-Y.} \bibnamefont{Wang}},
  \bibinfo{author}{\bibfnamefont{K.~W.} \bibnamefont{Edmonds}},
  \bibinfo{author}{\bibfnamefont{R.~P.} \bibnamefont{Campion}},
  \bibinfo{author}{\bibfnamefont{C.~R.} \bibnamefont{Staddon}},
  \bibinfo{author}{\bibfnamefont{N.~R.~S.} \bibnamefont{Farley}},
  \bibinfo{author}{\bibfnamefont{C.~T.} \bibnamefont{Foxon}},
  \bibinfo{author}{\bibfnamefont{E.}~\bibnamefont{Papis}},
  \bibinfo{author}{\bibfnamefont{E.}~\bibnamefont{Kami\ifmmode~\acute{n}\else
  \'{n}\fi{}ska}},
  \bibinfo{author}{\bibfnamefont{A.}~\bibnamefont{Piotrowska}},
  \bibnamefont{et~al.}, \bibinfo{journal}{Phys. Rev. B}
  \textbf{\bibinfo{volume}{71}}, \bibinfo{pages}{121302}
  (\bibinfo{year}{2005}).

\bibitem[{\citenamefont{Zemen et~al.}(2009)\citenamefont{Zemen, Ku\v{c}era,
  Olejn\'\i{}k, and Jungwirth}}]{Zemen:2009_PRB}
\bibinfo{author}{\bibfnamefont{J.}~\bibnamefont{Zemen}},
  \bibinfo{author}{\bibfnamefont{J.}~\bibnamefont{Ku\v{c}era}},
  \bibinfo{author}{\bibfnamefont{K.}~\bibnamefont{Olejn\'\i{}k}},
  \bibnamefont{and}
  \bibinfo{author}{\bibfnamefont{T.}~\bibnamefont{Jungwirth}},
  \bibinfo{journal}{Phys. Rev. B} \textbf{\bibinfo{volume}{80}},
  \bibinfo{pages}{155203} (\bibinfo{year}{2009}).

\end{thebibliography}
\end{document}